\documentclass[aps,prb,reprint,longbibliography,superscriptaddress,floatfix]{revtex4-2}

\usepackage{mathrsfs}
\usepackage{amsmath,gensymb}
\usepackage{amsfonts}
\usepackage{amssymb}
\usepackage{amsthm}
\usepackage{graphicx}
\usepackage{natbib}
\usepackage{hyperref}
\usepackage{bm}
\usepackage[caption=false]{subfig}
\usepackage{verbatim}

\graphicspath{{figures/}}

\hypersetup{hidelinks}

\begin{document}

\title{Multiple Andreev Reflection Spectroscopy of Spin-Resolved Proximity States}

\author{A.~A.~Golubov}
\affiliation{Moscow Institute of Physics and Technology, 141700 Dolgoprudny, Russia}
\affiliation{HSE University, 101000 Moscow, Russia}
\author{M.~Yu.~Kupriyanov}
\affiliation{Moscow Institute of Physics and Technology, 141700 Dolgoprudny, Russia}
\affiliation{Skobeltsyn Institute of Nuclear Physics, Lomonosov Moscow State University, Moscow 119991, Russian Federation}

\begin{abstract}
We study coherent multiple Andreev reflections (MAR) in short constrictions connecting superconductor--ferromagnet bilayers with parallel exchange fields. Proximity-induced spectral reconstruction produces split and redistributed differential-resistance anomalies, modifying both odd- and even-order subharmonic features, unlike a rigid Zeeman shift of BCS spectra. The shifted anomalies are approximately governed by the interval between the parent superconducting gap edge and an exchange-dependent spectral singularity. Qualitative spectral-interval estimates for antiparallel alignment are given in the Supplemental Material. Channel-transparency averaging extends the description from short ballistic constrictions to short diffusive multichannel junctions, where the characteristic anomalies survive. These findings provide a route to probing spin-resolved proximity states through two-terminal charge transport.
\end{abstract}

\date{}

\maketitle

\section{Introduction}
\label{sec:introduction}
Multiple Andreev reflections (MAR) in superconducting weak links generate the \begingroup subharmonic-gap structure (SGS)\endgroup{} at $eV=2\Delta/n$ and provide a sensitive spectroscopy of the quasiparticle spectrum through the energy-dependent Andreev amplitudes of the electrodes~\cite{Octavio1983}. \begingroup
Here we consider two bilayers, each formed by a superconductor (S) and an exchange-split conducting layer (F), joined by a short constriction ($c$). We show that, in this SF-$c$-FS (SFcFS) geometry\endgroup
, proximity-induced reconstruction of the spin-resolved electrode spectra produces split and redistributed MAR anomalies even for parallel exchange fields. The relevant scale is set not by a rigid displacement of \begingroup Bardeen--Cooper--Schrieffer (BCS)\endgroup{} gap edges, but by the interval between the outer gap edge and an exchange-induced \begingroup singularity in the density of states (DOS)\endgroup{}.

The microscopic theory of coherent MAR in short superconducting constrictions was developed by Gunsenheimer and Zaikin~\cite{Zaikin1994}, Bratus' et al.~\cite{Bratus1995}, Averin and Bardas~\cite{AverinBardas1995}, and Cuevas et al.~\cite{Cuevas1996} using different but equivalent theoretical approaches.
Subsequent work
established MAR spectroscopy in quantum point contacts, atomic-scale
junctions, semiconductor--superconductor devices, and related weak
links
\cite{Scheer1998,cron2001,Cuevas2006,Zaitsev1998,BrinkmanGolubov2000,Nilsson2012,Du2008,zhi2019,Yan2023}. Hybrid superconducting structures with magnetic
elements form a major platform of superconducting spintronics
\cite{golubovkup2004,Buzdin2005,bergeret2005,Eschrig2015,Birge2024}, while progress in
superconducting nanobridges and variable-thickness weak links with
magnetic layers or interfaces motivates the question of how
spin-resolved proximity spectra modify coherent MAR
\cite{batov2012,hoss2000multiple,krasnov2005planar,
batov2012double1096436,Krasnov2019,Golikova_2021,
ryazanov2025josephson}.

A common phenomenological interpretation of split SGS features invokes
an effective exchange field, \(eV_n\simeq 2(\Delta\pm h_{\rm eff})/n\).
This captures a kinematic shift of MAR thresholds, but assumes a
rigidly displaced BCS spectrum. MAR in superconductors with prescribed
Zeeman-split BCS spectra provides a useful benchmark
\cite{Lu2020}. For parallel spin
splittings in the two electrodes, the spectra are shifted in the same
way on both sides and the conventional MAR thresholds remain
unchanged. Splitting appears only for antiparallel or noncollinear spin
splittings: odd-order trajectories connect different electrodes and
probe relative spin-dependent gap-edge shifts, whereas even-order
trajectories return to the same electrode and remain pinned at the
usual subharmonic voltages \cite{Lu2020}.

Caldevilla-Asenjo \textit{et al.} recently observed exchange-induced
splitting of odd-order MAR peaks in EuS/Al/AlO$_x$/Al junctions with
one spin-split electrode~\cite{CaldevillaAsenjo2026}.
The absence of such splitting for even orders supports the
odd--even mechanism discussed by Lu \textit{et al.}~\cite{Lu2020}.
The proximity reconstruction considered here instead modifies both
odd- and even-order MAR features, even for parallel magnetizations.

Other mechanisms can also modify the SGS. Exchange-induced splitting
has been predicted for noncoherent MAR in long diffusive \begingroup
superconductor--insulator--ferromagnet--insulator--superconductor
(SIFIS) junctions\endgroup{}, where the effect relies on energy relaxation and
thermalization processes \cite{Polkin2023}. In magnetic quantum point
contacts, MAR anomalies may be modified by spin-active interface
scattering, spin filtering, and Andreev bound states
\cite{Bobkova2006,bobkova2007influence,PhysRevB.85.174510}. These mechanisms are distinct
from the proximity-induced spectral reconstruction considered here.

The theoretical framework used in this work connects the quasiclassical description of proximity-modified electrodes to the coherent \begingroup Averin--Bardas (AB)\endgroup{} MAR recurrence. The relation between Andreev reflection amplitudes and quasiclassical Green functions was established in Refs.~\cite{GolubovKupriyanov1995,Schopohl1998}. Earlier treatments of proximity-modified MAR transport include the \begingroup Octavio--Tinkham--Blonder--Klapwijk (OTBK)-based\endgroup{} analysis of Aminov \textit{et al.}~\cite{Aminov1996} and the coherent MAR formulation of Zaitsev and Averin~\cite{Zaitsev1998}. However, the consequences of spin-dependent proximity-induced spectral reconstruction for coherent MAR transport have not been investigated.

In the present formulation we introduce a local matrix Andreev coherence amplitude determined by the retarded quasiclassical Green functions of the electrodes. For arbitrary magnetic configurations the MAR recurrence is matrix-valued and side-dependent. Even collinear asymmetric junctions, including antiparallel SFcFS spin valves, require separate left and right spin-resolved amplitudes. 

Here we focus on the most transparent diagnostic regime of identical SF electrodes with parallel magnetizations and a spin-independent constriction. In this case the recurrence factorizes into two Nambu-spin sectors, but the spectral input in each sector is the proximity-reconstructed SF amplitude rather than a rigidly shifted BCS one. A rigid Zeeman-split BCS model with the same parallel alignment would not split the SGS. By contrast, we find that the proximity-reconstructed SF spectrum redistributes and splits the MAR anomalies. The modification of both odd- and even-order MAR features therefore reflects the internal spin-resolved spectral structure of the SF electrodes. \begingroup Antiparallel spin-valve geometries require separate left and right amplitudes within each sector and are left for future work.\endgroup{}

Earlier measurements on planar Al--(Cu/Fe)--Al bridges revealed split proximity-induced minigap features in the differential resistance, attributed to minigap splitting generated by the ferromagnetic layer~\cite{batov2012double1096436}. Related crosslike S--N/F--S structures demonstrated controllable Josephson transport under quasiparticle injection~\cite{Golikova_2021}. These experiments motivate coherent-MAR calculations for bridge-type geometries, where subharmonic-gap features can provide direct transport access to spin-resolved proximity spectra.

\begin{figure}
    \centering
    \includegraphics[width=\linewidth]{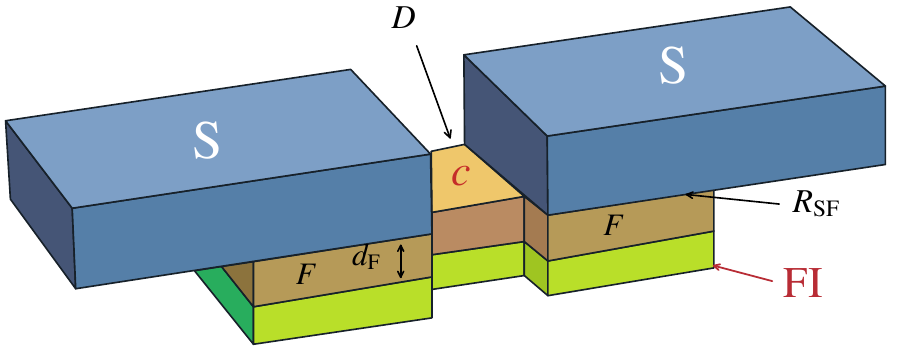}
    \caption{
\begingroup
Schematic of a possible SFcFS nanoconstriction. The conducting
spin-split \(F\) layers form the effective ferromagnetic parts of the
two SF electrodes. In one possible realization, these \(F\) layers are
initially normal-metal films whose spectra are exchange-split by the
adjacent ferromagnetic insulators (FI). The FI layers serve only as
sources of spin polarization and do not participate directly in
electronic transport through the junction. The short central connector
\(c\) forms the constriction and is characterized by the transmission
coefficient \(D\). The marked thickness \(d_F\) belongs to the
conducting F film, and \(R_{SF}\) is the specific resistance of SF interface, indicated by the arrow.
\endgroup
    }
    \label{fig:sample}
\end{figure}

\begingroup
A possible realization of our theoretical model is a variable-thickness
superconducting nanobridge incorporating a ferromagnetic insulator (FI),
as sketched in Fig.~\ref{fig:sample}. Here F denotes an effective
spin-split conducting layer: the brown film beneath each S bank may be
an initially normal metal whose spectrum is exchange-split by the
adjacent green FI~\cite{huertas2005proximity}. The FI supplies the
exchange field but carries no electronic current through the junction.
Each proximity-coupled S/F bilayer forms one composite SF electrode,
with the spectral functions evaluated in F next to the constriction.
The use of effective exchange fields in thin metallic S/F bilayers is
standard~\cite{Bergeret2001}; the fully averaged thin-S/thin-F limit in
that work is distinct from the rigid-S, thin-F spectrum used here.
For the illustrated realization, the FI is represented by the
prescribed homogeneous field $H$ in F, without resolving a separate
microscopic FI boundary problem. The central connector $c$ is short
and spin independent; spin-dependent scattering and phase accumulation
there are neglected. The schematic illustrates the effective model
rather than a new device architecture.

FI/S and FI/N structures are known to generate proximity-induced
exchange fields and spin-split spectra over a broad range of materials
and geometries~\cite{Hijano2021,hijano2022,Silaev2019,hao1990spin,hao1991thin,huertas2005proximity,PhysRevMaterials.1.054402,gomez2020strong}.
Related FI--S--FI spin-valve structures demonstrate that proximity-induced
exchange fields can become comparable to the superconducting energy
scale~\cite{hauser1969coupling,PhysRevLett.110.097001,zhu2017superconducting,golubov2017superconductivity43299613,NatCommun2025}.
More generally, the same effective SF-electrode description may also arise from other spin-injection or magnetic-proximity mechanisms, which makes the present analysis relevant beyond strictly FI-based structures.
Such extensions require the same equilibrium spectral-reservoir
approximation; injection-driven nonequilibrium distributions require a
separate treatment~\cite{Golikova_2021}.

Within this equilibrium setting, we ask how the reconstructed
spin-resolved electrode spectrum changes coherent MAR when the exchange
fields on the two sides are parallel. This separates the role of the
electrode spectrum from spin-dependent scattering at the connector.
Section~\ref{sec:spectral_transport} specifies the spectral input and
transport formulation, and Sec.~\ref{sec:results_discussion} examines the
resulting differential-resistance spectra and their spectroscopic
implications. Section~\ref{sec:summary} summarizes the conclusions.
Appendices~\ref{app:thinF_spectrum} and \ref{app:transport} give the
electrode derivation and the explicit transport prescription,
respectively.

\endgroup

\section{Spin-resolved spectral functions and MAR transport}
\label{sec:spectral_transport}

\begingroup
The spectral input to coherent MAR is the complex local Andreev
reflection amplitude at each electrode. Both its magnitude and phase
enter the MAR recurrence, so the local density of states alone is
insufficient. For the equilibrium electrodes and short, spin-independent
connector considered here, this amplitude is obtained from the local
retarded quasiclassical Green
functions~\cite{Zaitsev1998,GolubovKupriyanov1995,Schopohl1998}.
We first establish this relation and then specify the thin-F electrode
spectrum.

Let $u$ and $v$ denote the electron and hole spinors. We introduce the
time-reversed hole basis
$\bar v=i\hat\sigma_y v=(v_\downarrow,-v_\uparrow)^T$
and define the Andreev amplitude by $\bar v=\hat a u$.
Hats denote $2\times2$ spin matrices, and the superscript $R$ denotes
retarded quantities. The electrode phases are factored out locally;
their relative evolution is retained in the MAR formulation.

Let $\check g^R$ be the normalized $4\times4$ retarded quasiclassical
Green-function matrix in Nambu--spin space, with the physical solution
selected by the electrode boundary conditions. Since
$(\check g^R)^2=\check 1$, the matrix
$\check P_+=(\check 1+\check g^R)/2$ projects onto its $+1$ eigenspace.
In our convention, the upper-left and lower-left blocks of $\check g^R$
are $\hat G^R$ and $i\hat F^R$, respectively. Here $\hat G^R(E)$ and
$\hat F^R(E)$ are the dimensionless normal and anomalous retarded
quasiclassical Green functions, evaluated in the F film next to the
constriction. Applying $\check P_+$ to $(\chi,0)^T$, where $\chi$ is an
arbitrary coefficient spinor, gives
\[
u=\frac{\hat 1+\hat G^R}{2}\chi,
\qquad
\bar v=\frac{i\hat F^R}{2}\chi.
\]
The full Nambu-block convention and projector construction are given
in Sec.~II of the Supplementary Material~\cite{SupplementalMaterial}.

Eliminating $\chi$ gives
\[
\hat a(E)=i\hat F^R(E)
\bigl[\hat 1+\hat G^R(E)\bigr]^{-1}.
\]
The order of the factors follows from $\bar v=\hat a u$ and requires
no commutation assumption. Singular energies are understood by retarded
continuation. This local relation is independent of the thin-F
approximation used below. Here it supplies the electrode coherence
for equilibrium diffusive reservoirs, while normal scattering at the
short connector enters separately through the MAR recurrence.

For a common collinear exchange axis, singlet pairing couples
$e_\uparrow$ to $h_\downarrow$ in sector $+$ and $e_\downarrow$ to
$h_\uparrow$ in sector $-$. Spin-independent disorder, interfaces,
and connector scattering preserve these two sectors throughout the
MAR ladder. The local Green-function matrices are therefore diagonal
in the sector index. For the identical parallel electrodes considered
here, their left and right values coincide, giving the common
sector amplitudes
\begin{equation}
\label{eq:a_sigma_main}
a_\sigma(E)=\frac{iF_\sigma^R(E)}{1+G_\sigma^R(E)},
\qquad \sigma=\pm.
\end{equation}
The transport problem thus reduces to two independent scalar
Averin--Bardas recurrences, with generally different spectral inputs
for the two sectors. In a collinear asymmetric junction, the sectors
remain uncoupled, but separate left and right amplitudes are required.

For conventional BCS electrodes,
$G^R=z/Q$ and $F^R=-i\Delta/Q$, where $z=E+i0^+$ and
$Q=\sqrt{z^2-\Delta^2}$ has the retarded branch.
The resulting amplitude is the standard
$a_{\mathrm{BCS}}=\Delta/(z+Q)$.

\endgroup

\begin{figure}[!htbp]
    \centering
    \includegraphics[width=\linewidth]{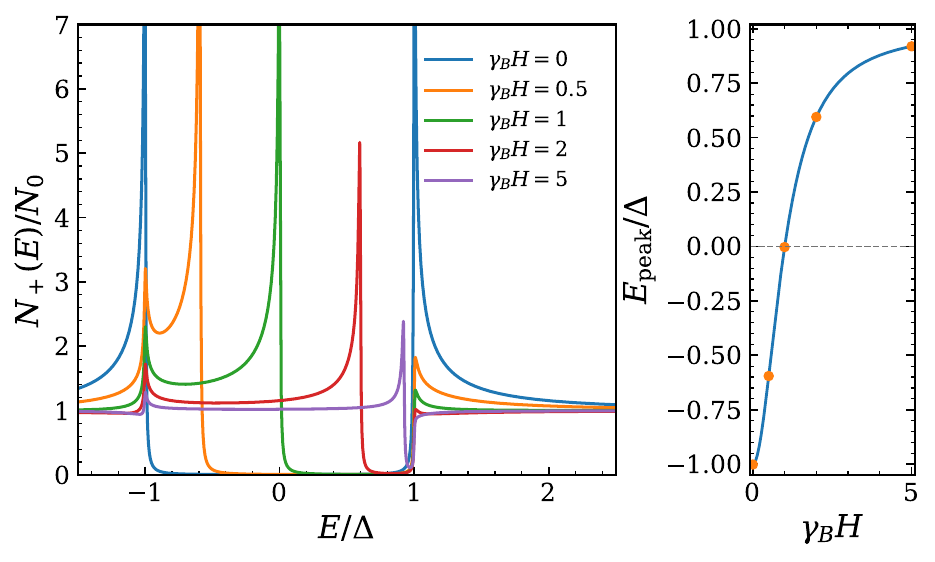}
    \caption{Local spin-resolved density of states
    \(N_+(E)=N_0{\rm Re}\,G_+^R(E)\) in the F-region of the SF electrode. The curves correspond to
    \(\gamma_BH=0,\,0.5,\,1,\,2,\) and \(5\), with
\(\Gamma=0.005\Delta\).
    With increasing \(\gamma_BH\), the positive-energy coherence
    peak remains close to \(E=\Delta\), while an exchange-induced
    spectral singularity shifts from \(E\simeq-\Delta\) toward
    higher energies. Right panel: position \(E_{\rm peak}\) of the shifted
    spectral peak as a function of \(\gamma_BH\).}
    \label{fig:DOS}
\end{figure}

\begingroup
As a concrete realization, we use the known thin-F limit of the
Usadel solution for a diffusive superconductor--ferromagnet (SF)
bilayer~\cite{Fominov2002}. We assume a homogeneous exchange field, a
spin-independent SF boundary, a rigid bulk BCS S reservoir, and
negligible spectral back-action of the small constriction. For this
equilibrium electrode spectrum, the retarded Green functions are
\endgroup
\begin{equation} \begin{aligned} G^R_\sigma(E) &= \frac{E_{\rm eff,\sigma}(E)} {Q_\sigma(E)}, \qquad F^R_\sigma(E) = -\frac{i\Delta}{Q_\sigma(E)}, \\ Q_\sigma(E) &= \sqrt{E_{\rm eff,\sigma}^2(E)-\Delta^2}. \end{aligned} \label{eq:GF_main} \end{equation}
The renormalized energy is
\begin{equation}
\begin{aligned}
 E_{\rm eff,\sigma}(E)&=E+i0^+\\
 &\quad+\frac{\gamma_B\left(E+i0^+-\sigma H\right)}{\pi T_c}
 \sqrt{\Delta^2-\left(E+i0^+\right)^2}.
\end{aligned}
\label{eq:Eeff_main}
\end{equation}
\begingroup
For numerical spectra, we replace $E+i0^+$ in Eq.~(3) consistently
by $z=E+i\Gamma$, including under the square root. Here $\Gamma>0$
is the spectral broadening parameter.
Equation~\eqref{eq:Eeff_main} is Eq.~(15) of
Ref.~\cite{Fominov2002}, written here for the two spin sectors. The new
result of the present work concerns the coherent-MAR response to this
proximity-reconstructed spectrum, rather than a new electrode solution.

The effective thin-F interface parameter is
\begin{equation}
 \gamma_B=\frac{R_{SF}}{\rho_F\xi_F}\frac{d_F}{\xi_F}
 =\frac{R_{SF}d_F}{\rho_F\xi_F^2}.
 \label{eq:gammaB_definition_main}
\end{equation}
Here \(\xi_F=\sqrt{\mathcal D_F/(2\pi T_c)}\) is the thermal diffusion
length, in units \(\hbar=k_B=1\). The quantities \(R_{SF}\), \(\rho_F\),
\(d_F\), and \(\mathcal D_F\) denote the specific resistance of SF interface, F-layer resistivity, thickness, and diffusion
constant, respectively. In Fig.~\ref{fig:sample}, $d_F$, $\rho_F$, and
$\xi_F$ refer to the brown conducting layer, whereas $R_{SF}$ refers to
the blue--brown interface, not to the F/FI boundary or to the
constriction. The thin-layer regime
requires \(d_F\ll\min(\xi_F,\sqrt{\mathcal D_F/(2|H|)})\) over the
relevant energy range. Appendix~\ref{app:thinF_spectrum} gives the
boundary conditions, thin-layer reduction and Matsubara-to-retarded
continuation. In the figure labels and discussion of numerical results,
\(\gamma_BH\) is a compact label for \(\gamma_B H/(\pi T_c)\);
\(H\) in Eq.~\eqref{eq:Eeff_main} is the dimensional exchange energy.
All numerical results presented here are evaluated at \(T=0\),
with \(\Delta\equiv\Delta(0)\) denoting the gap of the parent
superconductor. The temperature \(T\) is retained in the general
current expressions for completeness.

\endgroup
The spin-resolved density of states is 
\begin{equation} N_\sigma(E)=N_0{\rm Re}\,G^R_\sigma(E), \label{eq:DOS_main} 
\end{equation} 
where $N_0$ is the normal state density of states at the Fermi level.
\begingroup
Figure~\ref{fig:DOS} shows the positive-energy peak near $\Delta$ and
an exchange-induced peak moving from near $-\Delta$ toward positive
energies. Its signed position $E_{\rm peak}$ remains negative for the
MAR plots, where $E_p=|E_{\rm peak}|$. We use the small effective
parameter $\gamma_B=0.01$, for which the reconstruction is controlled
mainly by $\gamma_BH$. Since $\gamma_B$ includes $d_F/\xi_F$, this
value alone does not specify the microscopic interface transparency.
\endgroup

\begingroup
Pairing and exchange reside in different parts of the composite
electrode. The F film has no intrinsic pair potential: it acquires
electron--hole coherence from S, while $H$ changes the relative phase
in F. The rigid, nonmagnetic parent retains its gap $\Delta$, and its singular 
Green functions dominate the spectral boundary relation
near the parent edges (Appendix~\ref{app:thinF_spectrum}); away from
them, exchange reconstructs the induced spectrum.

At zero broadening, the correction in Eq.~\eqref{eq:Eeff_main}
vanishes at $E=\pm\Delta$. For $H>\Delta$, it is negative inside the
parent gap, excluding an internal root of $E_{\rm eff,+}=+\Delta$
but allowing an additional root $E_{\rm eff,+}=-\Delta$ at $E_s$.
This moving singularity is distinct from residual structure at
$-\Delta$.

Below the parent gap, S supports no propagating quasiparticles.
Andreev conversion at S/F and reflection at the outer film boundary
support subgap states confined normal to the film, with evanescent
tails in S~\cite{GolubovKupriyanov1995,Schopohl1998}. In the extended,
diffusive film, $E_s$ is a \emph{spectral edge of these states}, not an
isolated bound-state pole or a phase-dependent connector level. At a
simple edge, $G_+^R\propto(E-E_s)^{-1/2}$, whereas $a_+(E_s)=-1$
remains finite; its square-root energy dependence enters MAR.
For $H>\Delta$ and zero broadening,
\[
 N_+(E)=0,\qquad E_s<E<\Delta.
\]
This \emph{spin-resolved proximity gap} is an exchange-displaced
minigap~\cite{Fominov2002}, bounded by the moving edge $E_s$ and the
fixed parent edge. Near $+\Delta$, the singular S response dominates
the finite exchange contribution and leaves a nonempty gap. This is a
property of the present model, not a universal protection principle
for diffusive proximity systems.

The full Eq.~\eqref{eq:Eeff_main} verifies this restriction.
With $\eta=\gamma_B H/(\pi T_c)$, the internal root, distinct
from $E=-\Delta$, obeys
\[
 \eta=\frac{\gamma_B E_s}{\pi T_c}
      +\sqrt{\frac{\Delta+E_s}{\Delta-E_s}},
 \qquad -\Delta<E_s<\Delta.
\]
For $\gamma_B\ge0$, the right-hand side increases strictly from
$-\gamma_B\Delta/(\pi T_c)$ to $+\infty$ as $E_s$ traverses this
interval. Every finite positive $\eta$ therefore gives one root
below $\Delta$, crossing zero at $\eta=1$. For
$\gamma_B\Delta/(\pi T_c)\ll\min(1,\eta)$,
\[
 \frac{E_s}{\Delta}\simeq\frac{\eta^2-1}{\eta^2+1},\qquad
 \Delta-E_s\simeq\frac{2\Delta}{1+\eta^2}.
\]
The width remains positive at finite $\eta$: the moving edge
approaches $+\Delta$ only asymptotically. 

For $E>\Delta$, the retarded root is
$\sqrt{\Delta^2-(E+i0^+)^2}=-i\sqrt{E^2-\Delta^2}$, giving
\[
 E_{\rm eff,+}(E)=E-i\frac{\gamma_B(E-H)}{\pi T_c}
                         \sqrt{E^2-\Delta^2}.
\]
Its real part $E>\Delta$ excludes a real root
$E_{\rm eff,+}=\pm\Delta$. Propagating states in S provide an escape
channel: the subgap edge does not continue as a sharp above-gap
singularity, although broad continuum structure is possible.
The opposite sector obeys $N_-(E,H)=N_+(-E,H)$, so the spin-summed DOS
is even. Its common Fermi-level gap closes at $E_s=0$, while the
individual spin-resolved gaps persist. Finite $\Gamma$ softens these
gaps and can shift $E_{\rm peak}$ away from $E_s$ and the positive
peak away from $\Delta$. Coherent MAR thus samples the changing
interval between a parent edge and an internal proximity edge even
for identical, parallel electrodes.
\endgroup

The Averin--Bardas recurrence \cite{AverinBardas1995} is then used  with the BCS amplitude
at each MAR ladder rung replaced by the spin-resolved proximity
amplitude,
\begin{equation}
    a_{\rm BCS}(E_n)
    \longrightarrow
    a_\sigma(E_n,H),
    \qquad
    E_n=E+neV .
    \label{eq:AB_replacement_main}
\end{equation}

For the symmetric SFcFS junction considered here, the two SF
electrodes are identical and their magnetizations are parallel, while
the constriction is spin independent. Singlet Andreev reflection
converts \(e_\uparrow\leftrightarrow h_\downarrow\) within the
\(+\) sector and \(e_\downarrow\leftrightarrow h_\uparrow\) within
the \(-\) sector, but it does not mix the two sectors. \begingroup
The MAR problem therefore factorizes into two independent copies of
the scalar AB recurrence, one evaluated with \(a_+(E_n)\) and the other
with \(a_-(E_n)\).\endgroup{} In the standard AB notation
\(a_m=a(E+meV)\), this amounts to the replacement
\[
    a_m
    \rightarrow
    a_{\sigma,m}
    \equiv
    a_\sigma(E+meV,H),
    \qquad
    \sigma=\pm .
\]
\begingroup
The explicit recurrence relations, dc-current expression, and current
normalization are given in Appendix~\ref{app:transport}.\endgroup{}

We denote by \(I_\sigma(V,T;D)\) the dc current obtained from the
\begingroup corresponding sector recurrence\endgroup{}. These sector-resolved
contributions are not separately accessed in a conventional
two-terminal charge measurement. The observable current is obtained
only after summing the two Nambu--spin-sector contributions,
\begin{equation}
    I(V,T;D)
    =
    \frac{1}{2}
    \left[
        I_+(V,T;D)+I_-(V,T;D)
    \right].
    \label{eq:total_current_main}
\end{equation}
\begingroup
Appendix~\ref{app:transport} specifies the sector-current convention
and its bias-reversal symmetry, while numerical implementation details are provided in the Supplementary Material.\endgroup{}

\section{Results and discussion}
\label{sec:results_discussion}
We present the MAR spectra as differential resistance \(dV/dI\), the
representation commonly used in nanobridge experiments.
Figure~3 compares the conventional \begingroup
superconductor--constriction--superconductor (ScS)\endgroup{} reference with the
SFcFS response for \(\gamma_B=0.01\) and \(\gamma_BH=0.3\). For a
single channel with \(D=0.7\), the SFcFS spectrum is strongly
redistributed relative to the ScS reference.

\begin{figure}[!htbp]
    \centering
    \includegraphics[width=\linewidth]{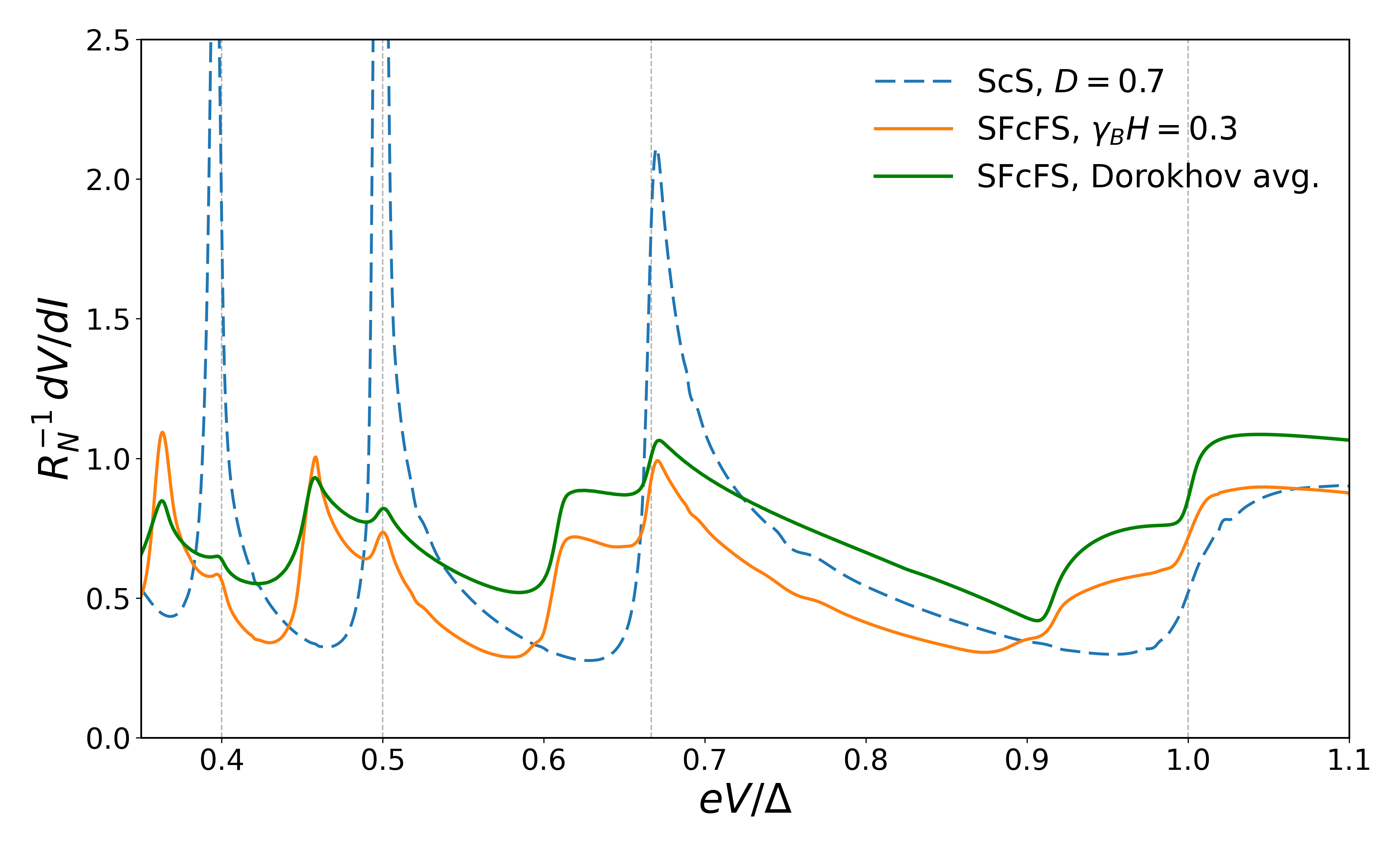}
\caption{
Reference MAR spectra in the differential-resistance representation.
The curves show a conventional ScS junction with \(D=0.7\), an SFcFS
junction with the same transparency, and the SFcFS response obtained by
averaging the current over the Dorokhov distribution of channel
transparencies. The SFcFS curves are
calculated for \(\gamma_B=0.01\), \(\gamma_BH=0.3\), and
\(\Gamma=0.005\Delta\). Vertical dashed lines mark the conventional BCS
MAR voltages \(eV/\Delta=2/n\). The exchange-induced subharmonic
anomalies remain visible after transmission averaging.
}
\label{fig:reference_transport}
\end{figure}

\begin{figure*}[t]
    \centering
    \includegraphics[width=1.0\textwidth]{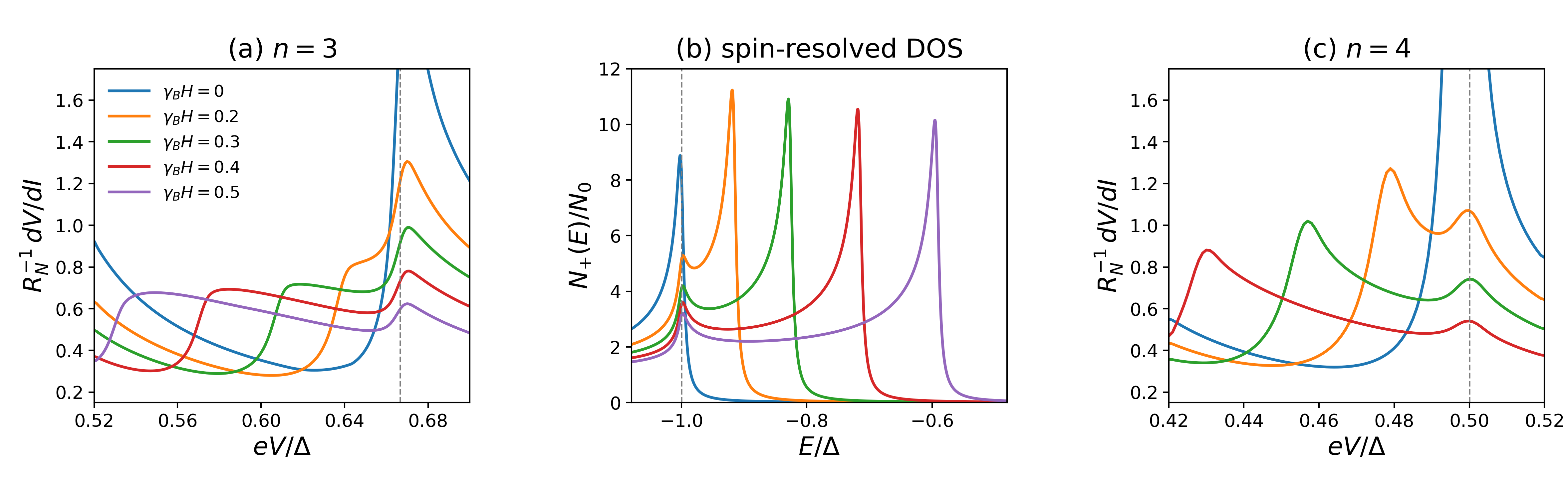}
    \caption{
   Evolution of MAR splitting and spin-resolved DOS in an SFcFS junction
with \(D=0.7\), \(\gamma_B=0.01\), and \(\Gamma=0.005\Delta\).
(a) \(R_N^{-1}dV/dI\) near the \(n=3\) anomaly for
\(\gamma_BH=0,0.2,0.3,0.4,0.5\). The dashed line marks
\(eV=2\Delta/3\). 
(b) Spin-resolved DOS \(N_+(E)/N_0\) for the same parameters; the dashed
line marks \(E=-\Delta\). 
(c) \(R_N^{-1}dV/dI\) near the \(n=4\) anomaly for
\(\gamma_BH=0,0.2,0.3,0.4\). The dashed line marks \(eV=\Delta/2\).
The field evolution shows that the shifted MAR component tracks the
proximity-induced spectral singularity. The modification of both
\(n=3\) and \(n=4\) anomalies confirms that the effect is governed by
the internal spin-resolved spectral structure of the SF electrodes.
}
    \label{fig:mar_splitting}
\end{figure*}

To model a short diffusive nanobridge with many conducting channels,
we average the single-channel current over the Dorokhov distribution
of channel transparencies,
\(\rho(D)\propto D^{-1}(1-D)^{-1/2}\)
\cite{dorokhov1984,Nazarov,Bardas1997,Beenakker1997}. The averaging is performed at
the level of the current before differentiating with respect to
voltage; \begingroup the weighting and normalization are specified in
Appendix~\ref{app:transport} 
and in Sec.~IV\,B
of the Supplementary Material.
\endgroup 
As shown in
Fig.~3, the sharp single-channel structures are broadened,
but the exchange-induced subharmonic anomalies remain clearly
resolved. The redistribution of the MAR structure therefore survives
transparency averaging and is a robust manifestation of
proximity-induced spectral reconstruction.

This behavior differs from a rigidly spin-shifted BCS spectrum. In
the parallel configuration considered here, such a rigid shift would
leave the conventional MAR thresholds at \(eV=2\Delta/n\). In the
SFcFS junction, by contrast, the exchange field reconstructs the local
spin-resolved spectral function itself. The MAR ladder therefore
probes a non-BCS spectrum containing both the outer coherence edge and
an additional exchange-induced singularity.

\begingroup
Figure~\ref{fig:mar_splitting} focuses on $n=3$ and $n=4$ as a
consecutive odd--even pair. In these windows, shifted peaks and shoulders
can be followed with increasing field and compared directly with the
same evolving spin-resolved electrode DOS. This provides a compact
illustration that proximity reconstruction modifies both parities in the
symmetric parallel configuration.
\par\endgroup

Figure~4 identifies this spectral origin directly. As
\(\gamma_BH\) increases, the negative-energy peak in \(N_+(E)\) moves
away from \(-\Delta\), and the low-voltage component of the split MAR
anomaly shifts accordingly. If the shifted singularity is located at
\(E=E_{\rm peak}<0\), the reconstructed spectral interval
\(\Delta+|E_{\rm peak}|\) gives the approximate MAR scale
\[
    eV_n^{(p)}
    \simeq
    \frac{\Delta+|E_{\rm peak}|}{n}.
\]
Equivalently, its separation from the conventional component at
\begingroup\(eV=2\Delta/n\) is controlled by
\[
    e\,\Delta V_n
    \simeq
    \frac{\Delta-|E_{\rm peak}|}{n}
    =
    \frac{\Delta+E_{\rm peak}}{n}.
\]
\endgroup
The simultaneous modification of the \(n=3\) and \(n=4\) anomalies
confirms that the effect is governed by the internal spin-resolved
spectral structure of the SF electrodes.

\begingroup
The choice of $n=3,4$ is illustrative: the full-range spectra in
Fig.~S1 also show exchange-dependent redistribution near
$eV/\Delta=2$ and $1$ ($n=1,2$). For the parameters shown, the
lower-order shifted contributions appear mainly as broad changes
of slope, making them less convenient for tracking individual
features. The selection is based on contrast against the background,
not an increased voltage separation, which decreases as $1/n$.

Coherent MAR probes the complex amplitude $a_\sigma(E)$, not the
DOS alone. The DOS peak marks the relevant spectral structure,
but the feature weights and line shapes also depend on coherent
scattering and channel transparency. These signatures remain visible
in the combined current of the two Nambu--spin sectors, allowing
spin-resolved proximity states to be probed in a conventional
two-terminal charge measurement.

The same framework could extend transmission-channel
characterization to magnetic proximity electrodes. For a
few-channel contact with normal electrodes,
the set of normal-state transparencies
$\{D_j\}$ can be inferred from MAR current--voltage
fitting~\cite{Scheer1998}. In the present symmetric parallel model,
a prospective fit would use $I(V)=\sum_j I_j(V;D_j,\eta)$, where
$I_j$ is the physical channel current including both spin sectors.
With the remaining spectral parameters specified, such a fit could
jointly constrain $\{D_j\}$ and $|\eta|$: spectral displacements
complement the transparency-dependent weights and line shapes,
while the normal-state conductance constrains $\sum_j D_j$.
The accuracy and uniqueness of this reconstruction remain to be
established. 

Figure~\ref{fig:reference_transport} shows that the exchange-induced
features also survive Dorokhov distribution averaging for the parameters
considered, connecting the single-channel mechanism to short
multichannel bridges. In the bridge-type structures motivating this
study~\cite{batov2012,batov2012double1096436,Golikova_2021}, a useful
experimental test would compare the field evolution of several MAR
orders with independent tunneling spectroscopy of the same
proximity electrodes. This would test the spectral-interval
interpretation rather than infer spin splitting from an isolated
transport peak. Within the equilibrium-reservoir approximation,
the charge-transport signature is determined by the local
conducting-electrode spectrum, not by whether an insulating or
metallic magnetic element supplies its exchange field.
\endgroup

\section{Summary}
\label{sec:summary}
We have developed a microscopic theory of coherent multiple Andreev
reflections in SFcFS nanoconstrictions and identified a MAR regime
controlled by spin-resolved proximity spectra. The exchange field
reconstructs the quasiparticle spectrum of the SF electrodes,
generating an exchange-induced spectral singularity that redistributes
and splits the subharmonic-gap anomalies.

These structures originate from proximity-induced reconstruction of the
spin-resolved SF spectrum, not from a rigid spin shift of BCS gap
edges, and remain visible after channel-transparency averaging.
Coherent MAR spectroscopy therefore provides a direct transport probe
of spin-resolved proximity states in non-BCS superconducting
electrodes.

The predicted effects should be accessible in variable-thickness
bridge-type structures, as sketched in Fig.~\ref{fig:sample}, whose
fabrication technology is well established
\cite{batov2012,batov2012double1096436,Golikova_2021,
ryazanov2025josephson}.

\begin{acknowledgments}
We acknowledge stimulating discussions with V.~V.~Ryazanov, T.~E.~Golikova and A.~Polkin. This work was supported by the Russian Science Foundation via the project No. 23-72-30004. 
\end{acknowledgments}

\appendix
\begingroup
\section{Thin-F electrode spectrum}
\label{app:thinF_spectrum}

We give the thin-F reduction of the Usadel boundary problem that leads to
Eq.~\eqref{eq:Eeff_main}. This is the known SF-bilayer result of
Ref.~\cite{Fominov2002}; the derivation also specifies the conventions
used here. We set $\hbar=k_B=1$ and factor out the phase of the parent
superconductor. The conducting F layer occupies $0<x<d_F$, with its S
interface at $x=0$. We use the equilibrium spectrum of the unperturbed
bilayer, neglecting the small constriction's spectral back-action.

The exchange field $H$ is homogeneous and collinear, and the F layer has
no intrinsic pair potential. Spin relaxation, orbital depairing, and
additional spin-dependent SF-interface scattering are neglected. The S
layer is a rigid bulk BCS reservoir with pair potential $\Delta$ and
critical temperature $T_c$. With $\mathcal D_F$ denoting the diffusion
constant, define
\begin{equation}
 \xi_F=\sqrt{\frac{\mathcal D_F}{2\pi T_c}},\qquad
 \gamma_B=\frac{R_{SF}d_F}{\rho_F\xi_F^2}.
 \label{eq:thinF_parameters_app}
\end{equation}
Here $R_{SF}$ is the specific resistance of SF interface and $\rho_F$
is the F-layer resistivity. A sufficient thin-layer regime
is $d_F\ll\min(\xi_F,\sqrt{\mathcal D_F/(2|H|)})$ over the relevant
energy range. At a specified higher Matsubara frequency, the additional
condition $|\widetilde\omega_\sigma|d_F^2/\mathcal D_F\ll1$ is
required. 

For positive Matsubara frequency $\omega$, introduce
$\widetilde\omega_\sigma=\omega+i\sigma H$, with $\sigma=\pm1$, and
$g_\sigma=\cos\theta_\sigma$, $f_\sigma=\sin\theta_\sigma$. The Usadel
equation in F is~\cite{Usadel1970,Fominov2002}
\begin{equation}
 \frac{\mathcal D_F}{2}\partial_x^2\theta_\sigma
 =\widetilde\omega_\sigma\sin\theta_\sigma.
 \label{eq:theta_usadel_app}
\end{equation}
With $x$ directed from S into F, the Kupriyanov--Lukichev boundary
condition and the reflecting outer-surface condition are~\cite{KupriyanovLukichev1988}
\begin{equation}
 \begin{aligned}
 \frac{R_{SF}}{\rho_F}\,\partial_x\theta_\sigma(0)
 &=\sin[\theta_\sigma(0)-\theta_S],\\
 \partial_x\theta_\sigma(d_F)&=0,
 \end{aligned}
 \label{eq:theta_boundaries_app}
\end{equation}
where
\begin{equation}
 \Omega_S=\sqrt{\omega^2+\Delta^2},\quad
 g_S=\frac{\omega}{\Omega_S},\quad
 f_S=\frac{\Delta}{\Omega_S}.
 \label{eq:theta_S_app}
\end{equation}
Integrating Eq.~\eqref{eq:theta_usadel_app} across the thin layer,
replacing its spectral functions by their leading, spatially uniform
values, and using Eq.~\eqref{eq:theta_boundaries_app}, gives
\begin{equation}
 \left(g_S+\frac{\gamma_B\widetilde\omega_\sigma}{\pi T_c}\right)
 f_\sigma=f_Sg_\sigma.
 \label{eq:thinF_integrated_app}
\end{equation}
No expansion in the anomalous amplitude is made. Defining
\begin{equation}
 W_\sigma(\omega)=\omega+
 \frac{\gamma_B\widetilde\omega_\sigma}{\pi T_c}\Omega_S,
 \label{eq:W_Matsubara_app}
\end{equation}
normalization yields
\begin{equation}
 g_\sigma=\frac{W_\sigma}{\sqrt{W_\sigma^2+\Delta^2}},\qquad
 f_\sigma=\frac{\Delta}{\sqrt{W_\sigma^2+\Delta^2}}.
 \label{eq:gf_Matsubara_app}
\end{equation}
The branch is selected continuously from positive real Matsubara
frequency. In the $\Phi$ parametrization of Ref.~\cite{Fominov2002},
$\Phi_{F,\sigma}=\widetilde\omega_\sigma f_\sigma/g_\sigma$, so
\begin{equation}
 \begin{aligned}
 \Phi_{F,\sigma}
 &=\frac{\widetilde\omega_\sigma\Delta}
 {\displaystyle\omega+
 \frac{\gamma_B\widetilde\omega_\sigma}{\pi T_c}\sqrt{\omega^2+\Delta^2}}\\
 &=\frac{(\widetilde\omega_\sigma/\omega)\Delta}
 {1+\gamma_B\widetilde\omega_\sigma/(\pi T_c g_S)}.
 \end{aligned}
 \label{eq:Phi_corrected_app}
\end{equation}

For the retarded functions, put $z=E+i0^+$ and continue
$\omega\mapsto-iz$. Then
$\widetilde\omega_\sigma\mapsto-i(z-\sigma H)$ and
$\Omega_S\mapsto s(z)=\sqrt{\Delta^2-z^2}$, with $s(i\omega)>0$.
Consequently,
\begin{equation}
 \begin{aligned}
 W_\sigma(-iz)&=-i\left[
 z+\frac{\gamma_B(z-\sigma H)}{\pi T_c}s(z)\right]\\
 &=-iE_{\mathrm{eff},\sigma}(z),
 \end{aligned}
 \label{eq:continuation_app}
\end{equation}
which is Eq.~\eqref{eq:Eeff_main}. The two square-root branches are
fixed by this continuation, not by independent principal-root choices.
Writing $Q_\sigma^2=E_{\mathrm{eff},\sigma}^2-\Delta^2$, with
$Q_\sigma(i\omega)=i\sqrt{W_\sigma(\omega)^2+\Delta^2}$, gives
$G_\sigma^R=E_{\mathrm{eff},\sigma}/Q_\sigma$.

The direct continuation of the auxiliary $f_\sigma=\sin\theta_\sigma$
is $f_\sigma^{\rm an}=i\Delta/Q_\sigma$. The anomalous-block convention
in the main text is $F_\sigma^R=-f_\sigma^{\rm an}$, hence
\begin{equation}
 \begin{aligned}
 F_\sigma^R&=-\frac{i\Delta}{Q_\sigma},\\
 a_\sigma&=\frac{iF_\sigma^R}{1+G_\sigma^R}
 =\frac{\Delta}{E_{\mathrm{eff},\sigma}+Q_\sigma}.
 \end{aligned}
 \label{eq:retarded_amplitude_app}
\end{equation}
This convention recovers the standard scalar BCS Andreev amplitude as
$\gamma_B\to0$ at fixed $H$. For the broadened numerical spectrum,
$0^+$ is replaced consistently by $\Gamma>0$ in both occurrences of $z$.

\par\endgroup

\begingroup
\section{Coherent MAR current and normalization}
\label{app:transport}

We present the transport formalism for the symmetric parallel junction. Its local
Andreev amplitudes are given by Eqs.~\eqref{eq:a_sigma_main}--\eqref{eq:Eeff_main}
and Appendix~\ref{app:thinF_spectrum}. The connector is short enough
that its normal scattering matrix is energy independent over the
relevant spectral range; spin filtering and spin mixing in the connector are neglected.
We calculate the time-averaged dc current at nonzero bias, rather than
the equilibrium Josephson current at a fixed phase.

\subsection{Scattering amplitudes and scalar recurrence}
\label{app:recurrence}
A convenient channel-phase convention is~\cite{AverinBardas1995,Bardas1997}
\begin{equation}
 S_e=\begin{pmatrix}\sqrt R&\sqrt D\\\sqrt D&-\sqrt R\end{pmatrix},
 \qquad S_h=S_e^*,\qquad R=1-D.
 \label{eq:scattering_appB}
\end{equation}
The matrices act on the left/right amplitudes and are proportional to
the identity in spin space. For one orbital channel with spin-independent
normal transmission, define
\begin{equation}
 R_Q=\frac{h}{2e^2},\qquad
 R_N(D)=\frac{R_Q}{D},\qquad e>0.
 \label{eq:resistances_appB}
\end{equation}
Thus the normal-state conductance of the physical channel, including
both sectors, is $2e^2D/h$.

At incident energy $E$, measured in the local reservoir frame, write
\begin{equation}
 \begin{aligned}
 a_{\sigma,m}&=a_\sigma(E+meV,H),\\
 w_\sigma(E)&=1-|a_{\sigma,0}|^2,\qquad
 J_\sigma(E)=\sqrt{w_\sigma(E)}.
 \end{aligned}
 \label{eq:source_appB}
\end{equation}
The source $J_\sigma$ describes injection from the equilibrium
reservoir. It must be counted only once in the current. To separate
this source weight from the recurrence, introduce unit-source amplitudes
$\mathcal A_{n,\sigma}$ and $\mathcal B_{n,\sigma}$ through
$A_{n,\sigma}=J_\sigma\mathcal A_{n,\sigma}$ and
$B_{n,\sigma}=J_\sigma\mathcal B_{n,\sigma}$.
Here $A$ and $B$ are the outgoing hole-like and reflected electron-like
amplitudes, respectively, on the left side of the connector. 

For compactness, let
\begin{equation}
 \begin{aligned}
 p_{n,\sigma}&=\frac{D a_{\sigma,2n+2}a_{\sigma,2n+1}}
                         {1-a_{\sigma,2n+1}^{\,2}},\\
 q_{n,\sigma}&=D\left(\frac{a_{\sigma,2n+1}^{\,2}}
                              {1-a_{\sigma,2n+1}^{\,2}}
                    +\frac{a_{\sigma,2n}^{\,2}}
                              {1-a_{\sigma,2n-1}^{\,2}}\right)
                   +1-a_{\sigma,2n}^{\,2},\\
 s_{n,\sigma}&=\frac{D a_{\sigma,2n}a_{\sigma,2n-1}}
                         {1-a_{\sigma,2n-1}^{\,2}}.
 \end{aligned}
 \label{eq:coefficients_appB}
\end{equation}
The scalar AB recurrence in the unit-source form of
Ref.~\cite{Bardas1997} is
\begin{equation}
 p_{n,\sigma}\mathcal B_{n+1,\sigma}
 -q_{n,\sigma}\mathcal B_{n,\sigma}
 +s_{n,\sigma}\mathcal B_{n-1,\sigma}
 =-\sqrt R\,\delta_{n0},
 \label{eq:B_recurrence_appB}
\end{equation}
followed by
\begin{equation}
 \begin{aligned}
 \mathcal A_{n+1,\sigma}
 &-a_{\sigma,2n+1}a_{\sigma,2n}\mathcal A_{n,\sigma}\\
 &=\sqrt R\left(a_{\sigma,2n+2}\mathcal B_{n+1,\sigma}
              -a_{\sigma,2n+1}\mathcal B_{n,\sigma}\right)
   +a_{\sigma,1}\delta_{n0}.
 \end{aligned}
 \label{eq:A_recurrence_appB}
\end{equation}
Multiplication by $J_\sigma$ gives the source-containing form used in
the Supplement. For positive bias, a finite ladder $-N\le n\le N$
can be solved with $\mathcal B_{-N,\sigma}=\mathcal B_{N,\sigma}=0$
and $\mathcal A_{-N,\sigma}=0$; $N$ is increased until endpoint effects
are negligible. The coefficients $a_{\sigma,m}$ are evaluated at all sideband
indices required by Eqs.~\eqref{eq:coefficients_appB}
and \eqref{eq:A_recurrence_appB}, including those immediately
outside the retained index range.

\subsection{Explicit dc current and sector normalization}
\label{app:current}
Define the real spectral kernel
\begin{equation}
 \begin{aligned}
 \mathcal K_\sigma(E,V;D)=w_\sigma(E)\Bigg\{&
 2\operatorname{Re}\!\left[a_{\sigma,0}\mathcal A_{0,\sigma}\right]\\
 {}+\sum_n(1+|a_{\sigma,2n}|^2)&
 \left(|\mathcal A_{n,\sigma}|^2
       -|\mathcal B_{n,\sigma}|^2\right)\Bigg\}.
 \end{aligned}
 \label{eq:kernel_appB}
\end{equation}
The first term is the interference of the incident electron with the
Andreev-generated electron amplitude. The sum collects the electron
and hole charge fluxes at the sidebands. In the source-containing
notation, the same kernel is
$2J_\sigma\operatorname{Re}(a_{\sigma,0}A_{0,\sigma})+
\sum_n(1+|a_{\sigma,2n}|^2)(|A_{n,\sigma}|^2-|B_{n,\sigma}|^2)$;
there is no extra factor $w_\sigma$ in this latter expression.

In the sector-current convention of
Eq.~\eqref{eq:total_current_main}, the explicit expression is
\begin{equation}
 \begin{aligned}
 I_\sigma=\frac{1}{eR_Q}\Bigg[DeV
 -\lim_{E_c\to\infty}\int_{-E_c}^{E_c}&dE\,
 \tanh\!\left(\frac{E}{2T}\right)\\[-2pt]
 &\times\mathcal K_\sigma(E,V;D)\Bigg].
 \end{aligned}
 \label{eq:current_appB}
\end{equation}
We set $k_B=1$; at $T=0$ the thermal factor is $\operatorname{sgn}E$.
This is the dc, symmetric-energy-cutoff form of Eq.~(11) of
Ref.~\cite{Bardas1997}, with the spin-dependent spectral input. The
explicit $DeV$ term is essential: it is the normal-current contribution
remaining after regularization of the filled-reservoir flux. 

For clarity, the incident-source part of the flux is
$w_\sigma+\mathcal K_\sigma$ before this regularization. Particle--hole
symmetry gives $w_+(E)=w_-(-E)$, so the spin-summed source weight is
even and its integral against $\tanh(E/2T)$ vanishes on symmetric
limits. This cancellation need not hold separately in each sector.
Equation~\eqref{eq:current_appB} specifies the bookkeeping convention
for $I_\sigma$; only the combination
$I=(I_++I_-)/2$ is the physical charge current. In the normal
limit, $a_{\sigma,m}=0$,
$\mathcal A_{n,\sigma}=0$, and
$\mathcal B_{n,\sigma}=\sqrt R\,\delta_{n0}$.
Then $\mathcal K_\sigma=-R$, its symmetric integral vanishes, and
Eq.~\eqref{eq:total_current_main} gives $I=V/R_N(D)$.

Equivalently, for $v=eV/\Delta$ and $x=E/\Delta$,
\begin{equation}
 \begin{aligned}
 j_\sigma(v;D)&\equiv\frac{eR_N(D)}{\Delta}I_\sigma(V,T;D)\\
 &=v-\frac{1}{D}\lim_{x_c\to\infty}\int_{-x_c}^{x_c}
 dx\,\tanh\!\left(\frac{\Delta x}{2T}\right)\\[-2pt]
 &\hspace{3.3em}\times\mathcal K_\sigma(\Delta x,V;D),\\
 j_D(v)&=\tfrac12[j_+(v;D)+j_-(v;D)].
 \end{aligned}
 \label{eq:dimensionless_current_appB}
\end{equation}
The identities
$a_\sigma(-E,H)=-a_{-\sigma}^*(E,H)$ imply
$I_\sigma(-V,H)=-I_{-\sigma}(V,H)$ and hence an odd physical current.
They also allow negative bias to be reconstructed from the positive-bias
calculation. As $\gamma_B\to0$ at fixed $H$, the two spectral inputs
coincide with the BCS amplitude and the physical current reduces to
the usual spin-degenerate AB result.

\subsection{Physical channel sum and transparency average}
\label{app:averaging}

We first sum the channel currents, then specialize to the Dorokhov
density for a diffusive connector and express the result in
dimensionless form.
Independent transmission channels carry current in parallel at the
same applied voltage $V$. Their charge currents therefore add,
with each $I(V;D_j)$ including both Nambu--spin sectors:
\begin{equation}
 I_{\rm tot}(V)=\sum_j I(V;D_j),\qquad
 \frac{1}{R_N^{\rm tot}}=\frac{1}{R_Q}\sum_j D_j.
 \label{eq:physical_sum_appB}
\end{equation}
Here $R_Q=h/(2e^2)$, $R_N(D)=R_Q/D$ is the normal resistance of one
channel, and $R_N^{\rm tot}$ is the resistance of their parallel
combination. For many channels, the sum can be expressed through the
transmission-eigenvalue density $\rho(D)$, where $\rho(D)dD$ counts
channels with transmission between $D$ and $D+dD$~\cite{Beenakker1997}:
\[
 \begin{aligned}
 I_{\rm tot}(V)&=\int_0^1 dD\,\rho(D)\,I(V;D),\\
 \frac{1}{R_N^{\rm tot}}&=\frac{1}{R_Q}
                         \int_0^1 dD\,D\rho(D).
 \end{aligned}
\]
Thus $\rho(D)$ is a channel density, not a probability distribution
normalized to unity. For a specified set of transmission eigenvalues,
the second relation determines the normal conductance.

For a short diffusive connector, we use the Dorokhov
density~\cite{dorokhov1984,Nazarov,Bardas1997},
\begin{equation}
 \rho(D)=\frac{R_Q}{2R_N^{\rm tot}}\frac{1}{D\sqrt{1-D}}.
 \label{eq:dorokhov_appB}
\end{equation}

Averaging with this distribution extends the single-channel
ballistic-constriction description to a short diffusive multichannel
connector. Elastic disorder is represented by the distribution of
transmission eigenvalues, while the electrode spectra and the
coherent MAR recurrence for each channel remain unchanged. The
averaging does not imply loss of phase coherence within a MAR
trajectory.

To express the channel sum in the dimensionless normalization of
Eq.~\eqref{eq:dimensionless_current_appB}, use $v=eV/\Delta$ and
\[
 I(V;D)=\frac{\Delta}{eR_N(D)}j_D(v)
       =\frac{\Delta D}{eR_Q}j_D(v).
\]
The current normalized to the total normal resistance is therefore
\begin{equation}
 \begin{aligned}
 j_{\rm av}(v)&\equiv\frac{eR_N^{\rm tot}}{\Delta}I_{\rm tot}(V)\\
 &=\frac{\displaystyle\int_0^1 dD\,\rho(D)D\,j_D(v)}
         {\displaystyle\int_0^1 dD\,\rho(D)D}.
 \end{aligned}
 \label{eq:weighted_average_appB}
\end{equation}
The factor $D$ arises solely from converting the channel-normalized
currents to physical units; it is not an additional weight in the
physical channel sum. The denominator accounts for normalization
by the total normal resistance. In the normal-state limit,
$j_D(v)=v$, so $j_{\rm av}(v)=v$ and
$I_{\rm tot}=V/R_N^{\rm tot}$.

For the Dorokhov density, the substitution \mbox{$u=\sqrt{1-D}$}
reduces Eq.~\eqref{eq:weighted_average_appB} to
\begin{equation}
 j_{\rm av}(v)=\int_0^1 du\,j_{1-u^2}(v),
 \label{eq:x_average_appB}
\end{equation}
which can be evaluated by quadrature without placing nodes exactly
at $D=0$ or $1$. The differential resistance is obtained from the
total current, not by averaging the individual channel resistances:
\begin{equation}
 \frac{1}{R_N^{\rm tot}}\frac{dV}{dI_{\rm tot}}
 =\left(\frac{dj_{\rm av}}{dv}\right)^{-1}.
 \label{eq:resistance_appB}
\end{equation}
For a single channel, the same relation holds with $j_D$ and $R_N(D)$.
Numerical implementation details are given in Sec.~IV\,B of the
Supplementary Material.

\par\endgroup

\onecolumngrid
\vspace*{18pt}
\twocolumngrid


\begin{thebibliography}{57}%
\makeatletter
\providecommand \@ifxundefined [1]{%
 \@ifx{#1\undefined}
}%
\providecommand \@ifnum [1]{%
 \ifnum #1\expandafter \@firstoftwo
 \else \expandafter \@secondoftwo
 \fi
}%
\providecommand \@ifx [1]{%
 \ifx #1\expandafter \@firstoftwo
 \else \expandafter \@secondoftwo
 \fi
}%
\providecommand \natexlab [1]{#1}%
\providecommand \enquote  [1]{``#1''}%
\providecommand \bibnamefont  [1]{#1}%
\providecommand \bibfnamefont [1]{#1}%
\providecommand \citenamefont [1]{#1}%
\providecommand \href@noop [0]{\@secondoftwo}%
\providecommand \href [0]{\begingroup \@sanitize@url \@href}%
\providecommand \@href[1]{\@@startlink{#1}\@@href}%
\providecommand \@@href[1]{\endgroup#1\@@endlink}%
\providecommand \@sanitize@url [0]{\catcode `\\12\catcode `\$12\catcode
  `\&12\catcode `\#12\catcode `\^12\catcode `\_12\catcode `\%12\relax}%
\providecommand \@@startlink[1]{}%
\providecommand \@@endlink[0]{}%
\providecommand \url  [0]{\begingroup\@sanitize@url \@url }%
\providecommand \@url [1]{\endgroup\@href {#1}{\urlprefix }}%
\providecommand \urlprefix  [0]{URL }%
\providecommand \Eprint [0]{\href }%
\providecommand \doibase [0]{https://doi.org/}%
\providecommand \selectlanguage [0]{\@gobble}%
\providecommand \bibinfo  [0]{\@secondoftwo}%
\providecommand \bibfield  [0]{\@secondoftwo}%
\providecommand \translation [1]{[#1]}%
\providecommand \BibitemOpen [0]{}%
\providecommand \bibitemStop [0]{}%
\providecommand \bibitemNoStop [0]{.\EOS\space}%
\providecommand \EOS [0]{\spacefactor3000\relax}%
\providecommand \BibitemShut  [1]{\csname bibitem#1\endcsname}%
\let\auto@bib@innerbib\@empty
\bibitem [{\citenamefont {Octavio}\ \emph {et~al.}(1983)\citenamefont
  {Octavio}, \citenamefont {Tinkham}, \citenamefont {Blonder},\ and\
  \citenamefont {Klapwijk}}]{Octavio1983}%
  \BibitemOpen
  \bibfield  {author} {\bibinfo {author} {\bibfnamefont {M.}~\bibnamefont
  {Octavio}}, \bibinfo {author} {\bibfnamefont {M.}~\bibnamefont {Tinkham}},
  \bibinfo {author} {\bibfnamefont {G.~E.}\ \bibnamefont {Blonder}},\ and\
  \bibinfo {author} {\bibfnamefont {T.~M.}\ \bibnamefont {Klapwijk}},\
  }\bibfield  {title} {\bibinfo {title} {Subharmonic energy-gap structure in
  superconducting constrictions},\ }\href
  {https://doi.org/10.1103/PhysRevB.27.6739} {\bibfield  {journal} {\bibinfo
  {journal} {Physical Review B}\ }\textbf {\bibinfo {volume} {27}},\ \bibinfo
  {pages} {6739} (\bibinfo {year} {1983})}\BibitemShut {NoStop}%
\bibitem [{\citenamefont {Gunsenheimer}\ and\ \citenamefont
  {Zaikin}(1994)}]{Zaikin1994}%
  \BibitemOpen
  \bibfield  {author} {\bibinfo {author} {\bibfnamefont {U.}~\bibnamefont
  {Gunsenheimer}}\ and\ \bibinfo {author} {\bibfnamefont {A.~D.}\ \bibnamefont
  {Zaikin}},\ }\bibfield  {title} {\bibinfo {title} {Ballistic charge transport
  in superconducting weak links},\ }\href
  {https://doi.org/10.1103/PhysRevB.50.6317} {\bibfield  {journal} {\bibinfo
  {journal} {Physical Review B}\ }\textbf {\bibinfo {volume} {50}},\ \bibinfo
  {pages} {6317} (\bibinfo {year} {1994})}\BibitemShut {NoStop}%
\bibitem [{\citenamefont {Bratus'}\ \emph {et~al.}(1995)\citenamefont
  {Bratus'}, \citenamefont {Shumeiko},\ and\ \citenamefont
  {Wendin}}]{Bratus1995}%
  \BibitemOpen
  \bibfield  {author} {\bibinfo {author} {\bibfnamefont {E.~N.}\ \bibnamefont
  {Bratus'}}, \bibinfo {author} {\bibfnamefont {V.~S.}\ \bibnamefont
  {Shumeiko}},\ and\ \bibinfo {author} {\bibfnamefont {G.}~\bibnamefont
  {Wendin}},\ }\bibfield  {title} {\bibinfo {title} {Theory of subharmonic gap
  structure in superconducting mesoscopic tunnel contacts},\ }\href
  {https://doi.org/10.1103/PhysRevLett.74.2110} {\bibfield  {journal} {\bibinfo
   {journal} {Physical Review Letters}\ }\textbf {\bibinfo {volume} {74}},\
  \bibinfo {pages} {2110} (\bibinfo {year} {1995})}\BibitemShut {NoStop}%
\bibitem [{\citenamefont {Averin}\ and\ \citenamefont
  {Bardas}(1995)}]{AverinBardas1995}%
  \BibitemOpen
  \bibfield  {author} {\bibinfo {author} {\bibfnamefont {D.}~\bibnamefont
  {Averin}}\ and\ \bibinfo {author} {\bibfnamefont {A.}~\bibnamefont
  {Bardas}},\ }\bibfield  {title} {\bibinfo {title} {ac {Josephson} effect in a
  single quantum channel},\ }\href
  {https://doi.org/10.1103/PhysRevLett.75.1831} {\bibfield  {journal} {\bibinfo
   {journal} {Physical Review Letters}\ }\textbf {\bibinfo {volume} {75}},\
  \bibinfo {pages} {1831} (\bibinfo {year} {1995})}\BibitemShut {NoStop}%
\bibitem [{\citenamefont {Cuevas}\ \emph {et~al.}(1996)\citenamefont {Cuevas},
  \citenamefont {Mart{\'{\i}}n-Rodero},\ and\ \citenamefont
  {Levy~Yeyati}}]{Cuevas1996}%
  \BibitemOpen
  \bibfield  {author} {\bibinfo {author} {\bibfnamefont {J.~C.}\ \bibnamefont
  {Cuevas}}, \bibinfo {author} {\bibfnamefont {A.}~\bibnamefont
  {Mart{\'{\i}}n-Rodero}},\ and\ \bibinfo {author} {\bibfnamefont {A.}~\bibnamefont
  {Levy~Yeyati}},\ }\bibfield  {title} {\bibinfo {title} {Hamiltonian approach
  to the transport properties of superconducting quantum point contacts},\
  }\href {https://doi.org/10.1103/PhysRevB.54.7366} {\bibfield  {journal}
  {\bibinfo  {journal} {Physical Review B}\ }\textbf {\bibinfo {volume} {54}},\
  \bibinfo {pages} {7366} (\bibinfo {year} {1996})}\BibitemShut {NoStop}%
\bibitem [{\citenamefont {Scheer}\ \emph {et~al.}(1998)\citenamefont {Scheer},
  \citenamefont {Agra{\"{\i}}t}, \citenamefont {Cuevas}, \citenamefont {Levy~Yeyati},
  \citenamefont {Ludoph}, \citenamefont {Mart{\'{\i}}n-Rodero}, \citenamefont
  {Rubio~Bollinger}, \citenamefont {van Ruitenbeek},\ and\ \citenamefont
  {Urbina}}]{Scheer1998}%
  \BibitemOpen
  \bibfield  {author} {\bibinfo {author} {\bibfnamefont {E.}~\bibnamefont
  {Scheer}}, \bibinfo {author} {\bibfnamefont {N.}~\bibnamefont {Agra{\"{\i}}t}},
  \bibinfo {author} {\bibfnamefont {J.~C.}\ \bibnamefont {Cuevas}}, \bibinfo
  {author} {\bibfnamefont {A.}~\bibnamefont {Levy~Yeyati}}, \bibinfo {author}
  {\bibfnamefont {B.}~\bibnamefont {Ludoph}}, \bibinfo {author} {\bibfnamefont
  {A.}~\bibnamefont {Mart{\'{\i}}n-Rodero}}, \bibinfo {author} {\bibfnamefont
  {G.}~\bibnamefont {Rubio~Bollinger}}, \bibinfo {author} {\bibfnamefont
  {J.~M.}\ \bibnamefont {van Ruitenbeek}},\ and\ \bibinfo {author}
  {\bibfnamefont {C.}~\bibnamefont {Urbina}},\ }\bibfield  {title} {\bibinfo
  {title} {The signature of chemical valence in the electrical conduction
  through a single-atom contact},\ }\href {https://doi.org/10.1038/28112}
  {\bibfield  {journal} {\bibinfo  {journal} {Nature}\ }\textbf {\bibinfo
  {volume} {394}},\ \bibinfo {pages} {154} (\bibinfo {year}
  {1998})}\BibitemShut {NoStop}%
\bibitem [{\citenamefont {Cron}\ \emph {et~al.}(2001)\citenamefont {Cron},
  \citenamefont {Goffman}, \citenamefont {Esteve},\ and\ \citenamefont
  {Urbina}}]{cron2001}%
  \BibitemOpen
  \bibfield  {author} {\bibinfo {author} {\bibfnamefont {R.}~\bibnamefont
  {Cron}}, \bibinfo {author} {\bibfnamefont {M.~F.}\ \bibnamefont {Goffman}},
  \bibinfo {author} {\bibfnamefont {D.}~\bibnamefont {Esteve}},\ and\ \bibinfo
  {author} {\bibfnamefont {C.}~\bibnamefont {Urbina}},\ }\bibfield  {title}
  {\bibinfo {title} {Multiple-charge-quanta shot noise in superconducting
  atomic contacts},\ }\href@noop {} {\bibfield  {journal} {\bibinfo  {journal}
  {Physical Review Letters}\ }\textbf {\bibinfo {volume} {86}},\ \bibinfo
  {pages} {4104} (\bibinfo {year} {2001})}\BibitemShut {NoStop}%
\bibitem [{\citenamefont {Cuevas}\ \emph {et~al.}(2006)\citenamefont {Cuevas},
  \citenamefont {Hammer}, \citenamefont {Kopu}, \citenamefont {Viljas},\ and\
  \citenamefont {Eschrig}}]{Cuevas2006}%
  \BibitemOpen
  \bibfield  {author} {\bibinfo {author} {\bibfnamefont {J.~C.}\ \bibnamefont
  {Cuevas}}, \bibinfo {author} {\bibfnamefont {J.}~\bibnamefont {Hammer}},
  \bibinfo {author} {\bibfnamefont {J.}~\bibnamefont {Kopu}}, \bibinfo {author}
  {\bibfnamefont {J.~K.}\ \bibnamefont {Viljas}},\ and\ \bibinfo {author}
  {\bibfnamefont {M.}~\bibnamefont {Eschrig}},\ }\bibfield  {title} {\bibinfo
  {title} {Proximity effect and multiple {Andreev} reflections in diffusive
  superconductor-normal-metal-superconductor junctions},\ }\href
  {https://doi.org/10.1103/PhysRevB.73.184505} {\bibfield  {journal} {\bibinfo
  {journal} {Phys. Rev. B}\ }\textbf {\bibinfo {volume} {73}},\ \bibinfo
  {pages} {184505} (\bibinfo {year} {2006})}\BibitemShut {NoStop}%
\bibitem [{\citenamefont {Zaitsev}\ and\ \citenamefont
  {Averin}(1998)}]{Zaitsev1998}%
  \BibitemOpen
  \bibfield  {author} {\bibinfo {author} {\bibfnamefont {A.~V.}\ \bibnamefont
  {Zaitsev}}\ and\ \bibinfo {author} {\bibfnamefont {D.~V.}\ \bibnamefont
  {Averin}},\ }\bibfield  {title} {\bibinfo {title} {Theory of ac {Josephson}
  effect in superconducting constrictions},\ }\href
  {https://doi.org/10.1103/PhysRevLett.80.3602} {\bibfield  {journal} {\bibinfo
   {journal} {Phys. Rev. Lett.}\ }\textbf {\bibinfo {volume} {80}},\ \bibinfo
  {pages} {3602} (\bibinfo {year} {1998})}\BibitemShut {NoStop}%
\bibitem [{\citenamefont {Brinkman}\ and\ \citenamefont
  {Golubov}(2000)}]{BrinkmanGolubov2000}%
  \BibitemOpen
  \bibfield  {author} {\bibinfo {author} {\bibfnamefont {A.}~\bibnamefont
  {Brinkman}}\ and\ \bibinfo {author} {\bibfnamefont {A.~A.}\ \bibnamefont
  {Golubov}},\ }\bibfield  {title} {\bibinfo {title} {Coherence effects in
  double-barrier {Josephson} junctions},\ }\href
  {https://doi.org/10.1103/PhysRevB.61.11297} {\bibfield  {journal} {\bibinfo
  {journal} {Physical Review B}\ }\textbf {\bibinfo {volume} {61}},\ \bibinfo
  {pages} {11297} (\bibinfo {year} {2000})}\BibitemShut {NoStop}%
\bibitem [{\citenamefont {Nilsson}\ \emph {et~al.}(2012)\citenamefont
  {Nilsson}, \citenamefont {Samuelsson}, \citenamefont {Caroff},\ and\
  \citenamefont {Xu}}]{Nilsson2012}%
  \BibitemOpen
  \bibfield  {author} {\bibinfo {author} {\bibfnamefont {H.~A.}\ \bibnamefont
  {Nilsson}}, \bibinfo {author} {\bibfnamefont {P.}~\bibnamefont {Samuelsson}},
  \bibinfo {author} {\bibfnamefont {P.}~\bibnamefont {Caroff}},\ and\ \bibinfo
  {author} {\bibfnamefont {H.~Q.}\ \bibnamefont {Xu}},\ }\bibfield  {title}
  {\bibinfo {title} {Supercurrent and multiple {Andreev} reflections in an
  {InSb} nanowire {Josephson} junction},\ }\href
  {https://doi.org/10.1021/nl203380w} {\bibfield  {journal} {\bibinfo
  {journal} {Nano Letters}\ }\textbf {\bibinfo {volume} {12}},\ \bibinfo
  {pages} {228} (\bibinfo {year} {2012})}\BibitemShut {NoStop}%
\bibitem [{\citenamefont {Du}\ \emph {et~al.}(2008)\citenamefont {Du},
  \citenamefont {Skachko},\ and\ \citenamefont {Andrei}}]{Du2008}%
  \BibitemOpen
  \bibfield  {author} {\bibinfo {author} {\bibfnamefont {X.}~\bibnamefont
  {Du}}, \bibinfo {author} {\bibfnamefont {I.}~\bibnamefont {Skachko}},\ and\
  \bibinfo {author} {\bibfnamefont {E.~Y.}\ \bibnamefont {Andrei}},\ }\bibfield
   {title} {\bibinfo {title} {Josephson current and multiple {Andreev}
  reflections in graphene {SNS} junctions},\ }\href
  {https://doi.org/10.1103/PhysRevB.77.184507} {\bibfield  {journal} {\bibinfo
  {journal} {Phys. Rev. B}\ }\textbf {\bibinfo {volume} {77}},\ \bibinfo
  {pages} {184507} (\bibinfo {year} {2008})}\BibitemShut {NoStop}%
\bibitem [{\citenamefont {Zhi}\ \emph {et~al.}(2019)\citenamefont {Zhi},
  \citenamefont {Kang}, \citenamefont {Li}, \citenamefont {Fan}, \citenamefont
  {Su}, \citenamefont {Pan}, \citenamefont {Zhao}, \citenamefont {Zhao},\ and\
  \citenamefont {Xu}}]{zhi2019}%
  \BibitemOpen
  \bibfield  {author} {\bibinfo {author} {\bibfnamefont {J.}~\bibnamefont
  {Zhi}}, \bibinfo {author} {\bibfnamefont {N.}~\bibnamefont {Kang}}, \bibinfo
  {author} {\bibfnamefont {S.}~\bibnamefont {Li}}, \bibinfo {author}
  {\bibfnamefont {D.}~\bibnamefont {Fan}}, \bibinfo {author} {\bibfnamefont
  {F.}~\bibnamefont {Su}}, \bibinfo {author} {\bibfnamefont {D.}~\bibnamefont
  {Pan}}, \bibinfo {author} {\bibfnamefont {S.}~\bibnamefont {Zhao}}, \bibinfo
  {author} {\bibfnamefont {J.}~\bibnamefont {Zhao}},\ and\ \bibinfo {author}
  {\bibfnamefont {H.}~\bibnamefont {Xu}},\ }\bibfield  {title} {\bibinfo
  {title} {Supercurrent and multiple {Andreev} reflections in {InSb} nanosheet
  {SNS} junctions},\ }\href@noop {} {\bibfield  {journal} {\bibinfo  {journal}
  {Physica Status Solidi (b)}\ }\textbf {\bibinfo {volume} {256}},\ \bibinfo
  {pages} {1800538} (\bibinfo {year} {2019})}\BibitemShut {NoStop}%
\bibitem [{\citenamefont {Yan}\ \emph {et~al.}(2023)\citenamefont {Yan},
  \citenamefont {Su}, \citenamefont {Pan}, \citenamefont {Li}, \citenamefont
  {Lyu}, \citenamefont {Chen}, \citenamefont {Wu}, \citenamefont {Lu},
  \citenamefont {Zhao}, \citenamefont {Wang},\ and\ \citenamefont
  {Xu}}]{Yan2023}%
  \BibitemOpen
  \bibfield  {author} {\bibinfo {author} {\bibfnamefont {S.}~\bibnamefont
  {Yan}}, \bibinfo {author} {\bibfnamefont {H.}~\bibnamefont {Su}}, \bibinfo
  {author} {\bibfnamefont {D.}~\bibnamefont {Pan}}, \bibinfo {author}
  {\bibfnamefont {W.}~\bibnamefont {Li}}, \bibinfo {author} {\bibfnamefont
  {Z.}~\bibnamefont {Lyu}}, \bibinfo {author} {\bibfnamefont {M.}~\bibnamefont
  {Chen}}, \bibinfo {author} {\bibfnamefont {X.}~\bibnamefont {Wu}}, \bibinfo
  {author} {\bibfnamefont {L.}~\bibnamefont {Lu}}, \bibinfo {author}
  {\bibfnamefont {J.}~\bibnamefont {Zhao}}, \bibinfo {author} {\bibfnamefont
  {J.-Y.}\ \bibnamefont {Wang}},\ and\ \bibinfo {author} {\bibfnamefont
  {H.}~\bibnamefont {Xu}},\ }\bibfield  {title} {\bibinfo {title}
  {Supercurrent, multiple {Andreev} reflections and {Shapiro} steps in {InAs}
  nanosheet {Josephson} junctions},\ }\href
  {https://doi.org/10.1021/acs.nanolett.3c01450} {\bibfield  {journal}
  {\bibinfo  {journal} {Nano Letters}\ }\textbf {\bibinfo {volume} {23}},\
  \bibinfo {pages} {6497} (\bibinfo {year} {2023})}\BibitemShut {NoStop}%
\bibitem [{\citenamefont {Golubov}\ \emph {et~al.}(2004)\citenamefont
  {Golubov}, \citenamefont {Kupriyanov},\ and\ \citenamefont
  {Il'ichev}}]{golubovkup2004}%
  \BibitemOpen
  \bibfield  {author} {\bibinfo {author} {\bibfnamefont {A.~A.}\ \bibnamefont
  {Golubov}}, \bibinfo {author} {\bibfnamefont {M.~Yu.}\ \bibnamefont
  {Kupriyanov}},\ and\ \bibinfo {author} {\bibfnamefont {E.}~\bibnamefont
  {Il'ichev}},\ }\bibfield  {title} {\bibinfo {title} {The current-phase
  relation in {Josephson} junctions},\ }\href@noop {} {\bibfield  {journal}
  {\bibinfo  {journal} {Reviews of Modern Physics}\ }\textbf {\bibinfo {volume}
  {76}},\ \bibinfo {pages} {411} (\bibinfo {year} {2004})}\BibitemShut
  {NoStop}%
\bibitem [{\citenamefont {Buzdin}(2005)}]{Buzdin2005}%
  \BibitemOpen
  \bibfield  {author} {\bibinfo {author} {\bibfnamefont {A.~I.}\ \bibnamefont
  {Buzdin}},\ }\bibfield  {title} {\bibinfo {title} {Proximity effects in
  superconductor-ferromagnet heterostructures},\ }\href
  {https://doi.org/10.1103/RevModPhys.77.935} {\bibfield  {journal} {\bibinfo
  {journal} {Rev. Mod. Phys.}\ }\textbf {\bibinfo {volume} {77}},\ \bibinfo
  {pages} {935} (\bibinfo {year} {2005})}\BibitemShut {NoStop}%
\bibitem [{\citenamefont {Bergeret}\ \emph {et~al.}(2005)\citenamefont
  {Bergeret}, \citenamefont {Volkov},\ and\ \citenamefont
  {Efetov}}]{bergeret2005}%
  \BibitemOpen
  \bibfield  {author} {\bibinfo {author} {\bibfnamefont {F.~S.}\ \bibnamefont
  {Bergeret}}, \bibinfo {author} {\bibfnamefont {A.~F.}\ \bibnamefont
  {Volkov}},\ and\ \bibinfo {author} {\bibfnamefont {K.~B.}\ \bibnamefont
  {Efetov}},\ }\bibfield  {title} {\bibinfo {title} {Odd triplet
  superconductivity and related phenomena in superconductor-ferromagnet
  structures},\ }\href@noop {} {\bibfield  {journal} {\bibinfo  {journal}
  {Reviews of Modern Physics}\ }\textbf {\bibinfo {volume} {77}},\ \bibinfo
  {pages} {1321} (\bibinfo {year} {2005})}\BibitemShut {NoStop}%
\bibitem [{\citenamefont {Eschrig}(2015)}]{Eschrig2015}%
  \BibitemOpen
  \bibfield  {author} {\bibinfo {author} {\bibfnamefont {M.}~\bibnamefont
  {Eschrig}},\ }\bibfield  {title} {\bibinfo {title} {Spin-polarized
  supercurrents for spintronics: a review of current progress},\ }\href
  {https://doi.org/10.1088/0034-4885/78/10/104501} {\bibfield  {journal}
  {\bibinfo  {journal} {Reports on Progress in Physics}\ }\textbf {\bibinfo
  {volume} {78}},\ \bibinfo {pages} {104501} (\bibinfo {year}
  {2015})}\BibitemShut {NoStop}%
\bibitem [{\citenamefont {Birge}\ and\ \citenamefont
  {Satchell}(2024)}]{Birge2024}%
  \BibitemOpen
  \bibfield  {author} {\bibinfo {author} {\bibfnamefont {N.~O.}\ \bibnamefont
  {Birge}}\ and\ \bibinfo {author} {\bibfnamefont {N.}~\bibnamefont
  {Satchell}},\ }\bibfield  {title} {\bibinfo {title} {Ferromagnetic materials
  for {Josephson} $\pi$ junctions},\ }\href {https://doi.org/10.1063/5.0195229}
  {\bibfield  {journal} {\bibinfo  {journal} {APL Materials}\ }\textbf
  {\bibinfo {volume} {12}},\ \bibinfo {pages} {041105} (\bibinfo {year}
  {2024})}\BibitemShut {NoStop}%
\bibitem [{\citenamefont {Golikova}\ \emph
  {et~al.}(2012{\natexlab{a}})\citenamefont {Golikova}, \citenamefont {H{\"u}bler},
  \citenamefont {Beckmann}, \citenamefont {Klenov}, \citenamefont {Bakurskiy},
  \citenamefont {Kupriyanov}, \citenamefont {Batov},\ and\ \citenamefont
  {Ryazanov}}]{batov2012}%
  \BibitemOpen
  \bibfield  {author} {\bibinfo {author} {\bibfnamefont {T.~E.}\ \bibnamefont
  {Golikova}}, \bibinfo {author} {\bibfnamefont {F.}~\bibnamefont {H{\"u}bler}},
  \bibinfo {author} {\bibfnamefont {D.}~\bibnamefont {Beckmann}}, \bibinfo
  {author} {\bibfnamefont {N.~V.}\ \bibnamefont {Klenov}}, \bibinfo {author}
  {\bibfnamefont {S.~V.}\ \bibnamefont {Bakurskiy}}, \bibinfo {author}
  {\bibfnamefont {M.~Yu.}\ \bibnamefont {Kupriyanov}}, \bibinfo {author}
  {\bibfnamefont {I.~E.}\ \bibnamefont {Batov}},\ and\ \bibinfo {author}
  {\bibfnamefont {V.~V.}\ \bibnamefont {Ryazanov}},\ }\bibfield  {title}
  {\bibinfo {title} {Critical current in planar {SNS} {Josephson} junctions},\
  }\href {https://doi.org/10.1134/S0021364012220043} {\bibfield  {journal}
  {\bibinfo  {journal} {Pis'ma ZhETP}\ }\textbf {\bibinfo {volume} {96}},\
  \bibinfo {pages} {743} (\bibinfo {year} {2012}{\natexlab{a}})},\ \bibinfo
  {note} {[JETP Letters \textbf{96} 668 (2012)]}\BibitemShut {NoStop}%
\bibitem [{\citenamefont {Hoss}\ \emph {et~al.}(2000)\citenamefont {Hoss},
  \citenamefont {Strunk}, \citenamefont {Nussbaumer}, \citenamefont {Huber},
  \citenamefont {Staufer},\ and\ \citenamefont
  {Sch{\"o}nenberger}}]{hoss2000multiple}%
  \BibitemOpen
  \bibfield  {author} {\bibinfo {author} {\bibfnamefont {T.}~\bibnamefont
  {Hoss}}, \bibinfo {author} {\bibfnamefont {C.}~\bibnamefont {Strunk}},
  \bibinfo {author} {\bibfnamefont {T.}~\bibnamefont {Nussbaumer}}, \bibinfo
  {author} {\bibfnamefont {R.}~\bibnamefont {Huber}}, \bibinfo {author}
  {\bibfnamefont {U.}~\bibnamefont {Staufer}},\ and\ \bibinfo {author}
  {\bibfnamefont {C.}~\bibnamefont {Sch{\"o}nenberger}},\ }\bibfield  {title}
  {\bibinfo {title} {Multiple {Andreev} reflection and giant excess noise in
  diffusive superconductor/normal-metal/superconductor junctions},\ }\href@noop
  {} {\bibfield  {journal} {\bibinfo  {journal} {Physical Review B}\ }\textbf
  {\bibinfo {volume} {62}},\ \bibinfo {pages} {4079} (\bibinfo {year}
  {2000})}\BibitemShut {NoStop}%
\bibitem [{\citenamefont {Krasnov}\ \emph {et~al.}(2005)\citenamefont
  {Krasnov}, \citenamefont {Ericsson}, \citenamefont {Intiso}, \citenamefont
  {Delsing}, \citenamefont {Oboznov}, \citenamefont {Prokofiev},\ and\
  \citenamefont {Ryazanov}}]{krasnov2005planar}%
  \BibitemOpen
  \bibfield  {author} {\bibinfo {author} {\bibfnamefont {V.~M.}\ \bibnamefont
  {Krasnov}}, \bibinfo {author} {\bibfnamefont {O.}~\bibnamefont {Ericsson}},
  \bibinfo {author} {\bibfnamefont {S.}~\bibnamefont {Intiso}}, \bibinfo
  {author} {\bibfnamefont {P.}~\bibnamefont {Delsing}}, \bibinfo {author}
  {\bibfnamefont {V.~A.}\ \bibnamefont {Oboznov}}, \bibinfo {author}
  {\bibfnamefont {A.~S.}\ \bibnamefont {Prokofiev}},\ and\ \bibinfo {author}
  {\bibfnamefont {V.~V.}\ \bibnamefont {Ryazanov}},\ }\bibfield  {title}
  {\bibinfo {title} {Planar {S--F--S Josephson} junctions made by focused ion
  beam etching},\ }\href@noop {} {\bibfield  {journal} {\bibinfo  {journal}
  {Physica C: Superconductivity}\ }\textbf {\bibinfo {volume} {418}},\ \bibinfo
  {pages} {16} (\bibinfo {year} {2005})}\BibitemShut {NoStop}%
\bibitem [{\citenamefont {Golikova}\ \emph
  {et~al.}(2012{\natexlab{b}})\citenamefont {Golikova}, \citenamefont
  {H{\"u}bler}, \citenamefont {Beckmann}, \citenamefont {Batov}, \citenamefont
  {Karminskaya}, \citenamefont {Kupriyanov}, \citenamefont {Golubov},\ and\
  \citenamefont {Ryazanov}}]{batov2012double1096436}%
  \BibitemOpen
  \bibfield  {author} {\bibinfo {author} {\bibfnamefont {T.~E.}\ \bibnamefont
  {Golikova}}, \bibinfo {author} {\bibfnamefont {F.}~\bibnamefont
  {H{\"u}bler}}, \bibinfo {author} {\bibfnamefont {D.}~\bibnamefont
  {Beckmann}}, \bibinfo {author} {\bibfnamefont {I.~E.}\ \bibnamefont {Batov}},
  \bibinfo {author} {\bibfnamefont {T.~Y.}\ \bibnamefont {Karminskaya}},
  \bibinfo {author} {\bibfnamefont {M.~Yu.}\ \bibnamefont {Kupriyanov}},
  \bibinfo {author} {\bibfnamefont {A.~A.}\ \bibnamefont {Golubov}},\ and\
  \bibinfo {author} {\bibfnamefont {V.~V.}\ \bibnamefont {Ryazanov}},\
  }\bibfield  {title} {\bibinfo {title} {Double proximity effect in hybrid
  planar superconductor-normal metal/ferromagnet-superconductor structures},\
  }\href {https://doi.org/10.1103/PhysRevB.86.064416} {\bibfield  {journal}
  {\bibinfo  {journal} {Physical Review B}\ }\textbf {\bibinfo {volume} {86}},\
  \bibinfo {pages} {064416} (\bibinfo {year} {2012}{\natexlab{b}})}\BibitemShut
  {NoStop}%
\bibitem [{\citenamefont {Golod}\ \emph {et~al.}(2019)\citenamefont {Golod},
  \citenamefont {Kapran},\ and\ \citenamefont {Krasnov}}]{Krasnov2019}%
  \BibitemOpen
  \bibfield  {author} {\bibinfo {author} {\bibfnamefont {T.}~\bibnamefont
  {Golod}}, \bibinfo {author} {\bibfnamefont {O.}~\bibnamefont {Kapran}},\ and\
  \bibinfo {author} {\bibfnamefont {V.}~\bibnamefont {Krasnov}},\ }\bibfield
  {title} {\bibinfo {title} {Planar superconductor-ferromagnet-superconductor
  {Josephson} junctions as scanning-probe sensors},\ }\href
  {https://doi.org/10.1103/PhysRevApplied.11.014062} {\bibfield  {journal}
  {\bibinfo  {journal} {Phys. Rev. Appl.}\ }\textbf {\bibinfo {volume} {11}},\
  \bibinfo {pages} {014062} (\bibinfo {year} {2019})}\BibitemShut {NoStop}%
\bibitem [{\citenamefont {Golikova}\ \emph {et~al.}(2021)\citenamefont
  {Golikova}, \citenamefont {Wolf}, \citenamefont {Beckmann}, \citenamefont
  {Penzyakov}, \citenamefont {Batov}, \citenamefont {Bobkova}, \citenamefont
  {Bobkov},\ and\ \citenamefont {Ryazanov}}]{Golikova_2021}%
  \BibitemOpen
  \bibfield  {author} {\bibinfo {author} {\bibfnamefont {T.~E.}\ \bibnamefont
  {Golikova}}, \bibinfo {author} {\bibfnamefont {M.~J.}\ \bibnamefont {Wolf}},
  \bibinfo {author} {\bibfnamefont {D.}~\bibnamefont {Beckmann}}, \bibinfo
  {author} {\bibfnamefont {G.~A.}\ \bibnamefont {Penzyakov}}, \bibinfo {author}
  {\bibfnamefont {I.~E.}\ \bibnamefont {Batov}}, \bibinfo {author}
  {\bibfnamefont {I.~V.}\ \bibnamefont {Bobkova}}, \bibinfo {author}
  {\bibfnamefont {A.~M.}\ \bibnamefont {Bobkov}},\ and\ \bibinfo {author}
  {\bibfnamefont {V.~V.}\ \bibnamefont {Ryazanov}},\ }\bibfield  {title}
  {\bibinfo {title} {Controllable supercurrent in mesoscopic
  superconductor-normal metal-ferromagnet crosslike {Josephson} structures},\
  }\href {https://doi.org/10.1088/1361-6668/abfd0d} {\bibfield  {journal}
  {\bibinfo  {journal} {Superconductor Science and Technology}\ }\textbf
  {\bibinfo {volume} {34}},\ \bibinfo {pages} {095001} (\bibinfo {year}
  {2021})}\BibitemShut {NoStop}%
\bibitem [{\citenamefont {Ryazanov}\ \emph {et~al.}(2025)\citenamefont
  {Ryazanov}, \citenamefont {Golikova}, \citenamefont {Bol'ginov},
  \citenamefont {Bobkova},\ and\ \citenamefont
  {Bobkov}}]{ryazanov2025josephson}%
  \BibitemOpen
  \bibfield  {author} {\bibinfo {author} {\bibfnamefont {V.~V.}\ \bibnamefont
  {Ryazanov}}, \bibinfo {author} {\bibfnamefont {T.~E.}\ \bibnamefont
  {Golikova}}, \bibinfo {author} {\bibfnamefont {V.~V.}\ \bibnamefont
  {Bol'ginov}}, \bibinfo {author} {\bibfnamefont {I.~V.}\ \bibnamefont
  {Bobkova}},\ and\ \bibinfo {author} {\bibfnamefont {A.~M.}\ \bibnamefont
  {Bobkov}},\ }\bibfield  {title} {\bibinfo {title} {Josephson $\pi $-junctions
  and {Andreev} current-transport 'engineering' in {S--N/F--S}
  nanostructures},\ }\href@noop {} {\bibfield  {journal} {\bibinfo  {journal}
  {Uspekhi Fizicheskikh Nauk}\ }\textbf {\bibinfo {volume} {195}},\ \bibinfo
  {pages} {766} (\bibinfo {year} {2025})},\ \bibinfo {note} {[Phys.-Uspekhi, \textbf{68} 717 (2025)]}\BibitemShut {NoStop}%
\bibitem [{\citenamefont {Lu}\ \emph {et~al.}(2020)\citenamefont {Lu},
  \citenamefont {Burset},\ and\ \citenamefont {Tanaka}}]{Lu2020}%
  \BibitemOpen
  \bibfield {author} {\bibinfo {author} {\bibfnamefont {B.}~\bibnamefont
  {Lu}}, \bibinfo {author} {\bibfnamefont {P.}~\bibnamefont {Burset}},\ and\
  \bibinfo {author} {\bibfnamefont {Y.}~\bibnamefont {Tanaka}},\ }
  \bibfield {title} {\bibinfo {title} {Spin-polarized multiple {Andreev}
  reflections in spin-split superconductors},\ }
  \href {https://doi.org/10.1103/PhysRevB.101.020502}
  {\bibfield {journal} {\bibinfo {journal} {Physical Review B}\ }\textbf
  {\bibinfo {volume} {101}},\ \bibinfo {pages} {020502(R)} (\bibinfo {year}
  {2020})}\BibitemShut {NoStop}%
\bibitem [{\citenamefont {Caldevilla-Asenjo}\ \emph {et~al.}(2026)}]{CaldevillaAsenjo2026}%
  \BibitemOpen
  \bibfield {author} {\bibinfo {author} {\bibfnamefont {D.}~\bibnamefont
  {Caldevilla-Asenjo}}, \bibinfo {author} {\bibfnamefont {G.~O.}\ \bibnamefont
  {Steffensen}}, \bibinfo {author} {\bibfnamefont {S.}~\bibnamefont
  {Catalano}}, \bibinfo {author} {\bibfnamefont {A.}~\bibnamefont
  {Hijano}}, \bibinfo {author} {\bibfnamefont {M.}~\bibnamefont
  {Ilyn}}, \bibinfo {author} {\bibfnamefont {C.}~\bibnamefont
  {Rogero}}, \bibinfo {author} {\bibfnamefont {R.}~\bibnamefont
  {Aguado}}, \bibinfo {author} {\bibfnamefont {F.~S.}\ \bibnamefont
  {Bergeret}},\ and\ \bibinfo {author} {\bibfnamefont {A.}~\bibnamefont
  {Levy~Yeyati}},\ }
  \bibfield {title} {\bibinfo {title} {Coherent subgap transport in
  spin-split {Josephson} junctions},\ }
  \href {https://doi.org/10.1103/lt7q-6ff3}
  {\bibfield {journal} {\bibinfo {journal} {Phys. Rev. Lett.}\ }\textbf
  {\bibinfo {volume} {136}},\ \bibinfo {pages} {176301} (\bibinfo {year}
  {2026})}\BibitemShut {NoStop}%
\bibitem [{\citenamefont {Polkin}\ and\ \citenamefont
  {Ioselevich}(2023)}]{Polkin2023}%
  \BibitemOpen
  \bibfield  {author} {\bibinfo {author} {\bibfnamefont {A.~V.}\ \bibnamefont
  {Polkin}}\ and\ \bibinfo {author} {\bibfnamefont {P.~A.}\ \bibnamefont
  {Ioselevich}},\ }\bibfield  {title} {\bibinfo {title} {Multiple {Andreev}
  reflections in diffusive {SINIS} and {SIFIS} junctions},\ }\href
  {https://doi.org/10.21468/SciPostPhys.15.3.100} {\bibfield  {journal}
  {\bibinfo  {journal} {SciPost Physics}\ }\textbf {\bibinfo {volume} {15}},\
  \bibinfo {pages} {100} (\bibinfo {year} {2023})}\BibitemShut {NoStop}%
\bibitem [{\citenamefont {Bobkova}(2006)}]{Bobkova2006}%
  \BibitemOpen
  \bibfield  {author} {\bibinfo {author} {\bibfnamefont {I.~V.}\ \bibnamefont
  {Bobkova}},\ }\bibfield  {title} {\bibinfo {title} {Subharmonic gap structure
  in superconductor/ferromagnet/superconductor junctions},\ }\href
  {https://doi.org/10.1103/PhysRevB.73.012506} {\bibfield  {journal} {\bibinfo
  {journal} {Phys. Rev. B}\ }\textbf {\bibinfo {volume} {73}},\ \bibinfo
  {pages} {012506} (\bibinfo {year} {2006})}\BibitemShut {NoStop}%
\bibitem [{\citenamefont {Bobkova}\ and\ \citenamefont
  {Bobkov}(2007)}]{bobkova2007influence}%
  \BibitemOpen
  \bibfield  {author} {\bibinfo {author} {\bibfnamefont {I.}~\bibnamefont
  {Bobkova}}\ and\ \bibinfo {author} {\bibfnamefont {A.}~\bibnamefont
  {Bobkov}},\ }\bibfield  {title} {\bibinfo {title} {Influence of spin
  filtering and spin mixing on the subgap structure of {I-V} characteristics in
  a superconducting quantum point contact},\ }\href@noop {} {\bibfield
  {journal} {\bibinfo  {journal} {Physical Review B---Condensed Matter and
  Materials Physics}\ }\textbf {\bibinfo {volume} {76}},\ \bibinfo {pages}
  {094517} (\bibinfo {year} {2007})}\BibitemShut {NoStop}%
\bibitem [{\citenamefont {Popovi{\'c}}\ \emph {et~al.}(2012)\citenamefont
  {Popovi{\'c}}, \citenamefont {Dobrosavljevi{\'c}-Gruji{\'c}},\ and\
  \citenamefont {Zikic}}]{PhysRevB.85.174510}%
  \BibitemOpen
  \bibfield  {author} {\bibinfo {author} {\bibfnamefont {Z.}~\bibnamefont
  {Popovi{\'c}}}, \bibinfo {author} {\bibfnamefont {L.}~\bibnamefont
  {Dobrosavljevi{\'c}-Gruji{\'c}}},\ and\ \bibinfo {author} {\bibfnamefont
  {R.}~\bibnamefont {Zikic}},\ }\bibfield  {title} {\bibinfo {title} {Influence
  of {Andreev} reflection on current-voltage characteristics of
  superconductor/ferromagnet/superconductor metallic weak links},\ }\href@noop
  {} {\bibfield  {journal} {\bibinfo  {journal} {Physical Review B---Condensed
  Matter and Materials Physics}\ }\textbf {\bibinfo {volume} {85}},\ \bibinfo
  {pages} {174510} (\bibinfo {year} {2012})}\BibitemShut {NoStop}%
\bibitem [{\citenamefont {Golubov}\ and\ \citenamefont
  {Kupriyanov}(1995)}]{GolubovKupriyanov1995}%
  \BibitemOpen
  \bibfield  {author} {\bibinfo {author} {\bibfnamefont {A.~A.}\ \bibnamefont
  {Golubov}}\ and\ \bibinfo {author} {\bibfnamefont {M.~Yu.}\ \bibnamefont
  {Kupriyanov}},\ }\bibfield  {title} {\bibinfo {title} {Quasiparticle current
  of ballistic {NcS'S} contacts},\ }\href@noop {} {\bibfield  {journal}
  {\bibinfo  {journal} {Pis'ma ZhETP}\ }\textbf {\bibinfo {volume} {61}},\
  \bibinfo {pages} {830} (\bibinfo {year} {1995})},\ \bibinfo {note} {[JETP
  Letters, \textbf{61} 851 (1995)]}\BibitemShut {NoStop}%
\bibitem [{\citenamefont {Schopohl}()}]{Schopohl1998}%
  \BibitemOpen
  \bibfield  {author} {\bibinfo {author} {\bibfnamefont {N.}~\bibnamefont
  {Schopohl}},\ }\href@noop {} {\bibinfo {title} {Transformation of the
  {Eilenberger} equations of superconductivity to a scalar {Riccati}
  equation}},\ \Eprint {https://arxiv.org/abs/cond-mat/9804064}
  {arXiv:cond-mat/9804064 [cond-mat.supr-con]} \BibitemShut {NoStop}%
\bibitem [{\citenamefont {Aminov}\ \emph {et~al.}(1996)\citenamefont {Aminov},
  \citenamefont {Golubov},\ and\ \citenamefont {Kupriyanov}}]{Aminov1996}%
  \BibitemOpen
  \bibfield  {author} {\bibinfo {author} {\bibfnamefont {B.~A.}\ \bibnamefont
  {Aminov}}, \bibinfo {author} {\bibfnamefont {A.~A.}\ \bibnamefont
  {Golubov}},\ and\ \bibinfo {author} {\bibfnamefont {M.~Yu.}\ \bibnamefont
  {Kupriyanov}},\ }\bibfield  {title} {\bibinfo {title} {Quasiparticle current
  in ballistic constrictions with finite transparencies of interfaces},\ }\href
  {https://doi.org/10.1103/PhysRevB.53.365} {\bibfield  {journal} {\bibinfo
  {journal} {Physical Review B}\ }\textbf {\bibinfo {volume} {53}},\ \bibinfo
  {pages} {365} (\bibinfo {year} {1996})}\BibitemShut {NoStop}%
\bibitem [{\citenamefont {Huertas-Hernando}\ and\ \citenamefont
  {Nazarov}(2005)}]{huertas2005proximity}%
  \BibitemOpen
  \bibfield  {author} {\bibinfo {author} {\bibfnamefont {D.}~\bibnamefont
  {Huertas-Hernando}}\ and\ \bibinfo {author} {\bibfnamefont {Y.~V.}\
  \bibnamefont {Nazarov}},\ }\bibfield  {title} {\bibinfo {title} {Proximity
  effect gaps in {S/N/FI} structures},\ }\href@noop {} {\bibfield  {journal}
  {\bibinfo  {journal} {The European Physical Journal B-Condensed Matter and
  Complex Systems}\ }\textbf {\bibinfo {volume} {44}},\ \bibinfo {pages} {373}
  (\bibinfo {year} {2005})}\BibitemShut {NoStop}%
\bibitem [{\citenamefont {Bergeret}\ \emph {et~al.}(2001)\citenamefont
  {Bergeret}, \citenamefont {Volkov},\ and\ \citenamefont
  {Efetov}}]{Bergeret2001}%
  \BibitemOpen
  \bibfield  {author} {\bibinfo {author} {\bibfnamefont {F.~S.}\ \bibnamefont
  {Bergeret}}, \bibinfo {author} {\bibfnamefont {A.~F.}\ \bibnamefont
  {Volkov}},\ and\ \bibinfo {author} {\bibfnamefont {K.~B.}\ \bibnamefont
  {Efetov}},\ }\bibfield  {title} {\bibinfo {title} {Enhancement of the
  {Josephson} current by an exchange field in superconductor-ferromagnet
  structures},\ }\href {https://doi.org/10.1103/PhysRevLett.86.3140} {\bibfield
   {journal} {\bibinfo  {journal} {Phys. Rev. Lett.}\ }\textbf {\bibinfo
  {volume} {86}},\ \bibinfo {pages} {3140} (\bibinfo {year}
  {2001})}\BibitemShut {NoStop}%
\bibitem [{\citenamefont {Hijano}\ \emph {et~al.}(2021)\citenamefont {Hijano},
  \citenamefont {Ili\ifmmode~\acute{c}\else \'{c}\fi{}}, \citenamefont {Rouco},
  \citenamefont {Gonz\'alez-Orellana}, \citenamefont {Ilyn}, \citenamefont
  {Rogero}, \citenamefont {Virtanen}, \citenamefont {Heikkil\"a}, \citenamefont
  {Khorshidian}, \citenamefont {Spies}, \citenamefont {Ligato}, \citenamefont
  {Giazotto}, \citenamefont {Strambini},\ and\ \citenamefont
  {Bergeret}}]{Hijano2021}%
  \BibitemOpen
  \bibfield  {author} {\bibinfo {author} {\bibfnamefont {A.}~\bibnamefont
  {Hijano}}, \bibinfo {author} {\bibfnamefont {S.}~\bibnamefont
  {Ili\ifmmode~\acute{c}\else \'{c}\fi{}}}, \bibinfo {author} {\bibfnamefont
  {M.}~\bibnamefont {Rouco}}, \bibinfo {author} {\bibfnamefont
  {C.}~\bibnamefont {Gonz\'alez-Orellana}}, \bibinfo {author} {\bibfnamefont
  {M.}~\bibnamefont {Ilyn}}, \bibinfo {author} {\bibfnamefont {C.}~\bibnamefont
  {Rogero}}, \bibinfo {author} {\bibfnamefont {P.}~\bibnamefont {Virtanen}},
  \bibinfo {author} {\bibfnamefont {T.~T.}\ \bibnamefont {Heikkil\"a}},
  \bibinfo {author} {\bibfnamefont {S.}~\bibnamefont {Khorshidian}}, \bibinfo
  {author} {\bibfnamefont {M.}~\bibnamefont {Spies}}, \bibinfo {author}
  {\bibfnamefont {N.}~\bibnamefont {Ligato}}, \bibinfo {author} {\bibfnamefont
  {F.}~\bibnamefont {Giazotto}}, \bibinfo {author} {\bibfnamefont
  {E.}~\bibnamefont {Strambini}},\ and\ \bibinfo {author} {\bibfnamefont
  {F.~S.}\ \bibnamefont {Bergeret}},\ }\bibfield  {title} {\bibinfo {title}
  {Coexistence of superconductivity and spin-splitting fields in
  superconductor/ferromagnetic insulator bilayers of arbitrary thickness},\
  }\href {https://doi.org/10.1103/PhysRevResearch.3.023131} {\bibfield
  {journal} {\bibinfo  {journal} {Phys. Rev. Research}\ }\textbf {\bibinfo
  {volume} {3}},\ \bibinfo {pages} {023131} (\bibinfo {year}
  {2021})}\BibitemShut {NoStop}%
\bibitem [{\citenamefont {Hijano}\ \emph {et~al.}(2022)\citenamefont {Hijano},
  \citenamefont {Golovach},\ and\ \citenamefont {Bergeret}}]{hijano2022}%
  \BibitemOpen
  \bibfield  {author} {\bibinfo {author} {\bibfnamefont {A.}~\bibnamefont
  {Hijano}}, \bibinfo {author} {\bibfnamefont {V.~N.}\ \bibnamefont
  {Golovach}},\ and\ \bibinfo {author} {\bibfnamefont {F.~S.}\ \bibnamefont
  {Bergeret}},\ }\bibfield  {title} {\bibinfo {title} {Quasiparticle density of
  states and triplet correlations in superconductor/ferromagnetic-insulator
  structures across a sharp domain wall},\ }\href@noop {} {\bibfield  {journal}
  {\bibinfo  {journal} {Physical Review B}\ }\textbf {\bibinfo {volume}
  {105}},\ \bibinfo {pages} {174507} (\bibinfo {year} {2022})}\BibitemShut
  {NoStop}%
\bibitem [{\citenamefont {Heikkil{\"a}}\ \emph {et~al.}(2019)\citenamefont
  {Heikkil{\"a}}, \citenamefont {Silaev}, \citenamefont {Virtanen},\ and\
  \citenamefont {Bergeret}}]{Silaev2019}%
  \BibitemOpen
  \bibfield  {author} {\bibinfo {author} {\bibfnamefont {T.~T.}\ \bibnamefont
  {Heikkil{\"a}}}, \bibinfo {author} {\bibfnamefont {M.}~\bibnamefont
  {Silaev}}, \bibinfo {author} {\bibfnamefont {P.}~\bibnamefont {Virtanen}},\
  and\ \bibinfo {author} {\bibfnamefont {F.~S.}\ \bibnamefont {Bergeret}},\
  }\bibfield  {title} {\bibinfo {title} {Thermal, electric and spin transport
  in superconductor/ferromagnetic-insulator structures},\ }\href
  {https://doi.org/10.1016/j.progsurf.2019.100540} {\bibfield
  {journal} {\bibinfo  {journal} {Progress in Surface Science}\ }\textbf
  {\bibinfo {volume} {94}},\ \bibinfo {pages} {100540} (\bibinfo {year}
  {2019})}\BibitemShut {NoStop}%
\bibitem [{\citenamefont {Hao}\ \emph {et~al.}(1990)\citenamefont {Hao},
  \citenamefont {Moodera},\ and\ \citenamefont {Meservey}}]{hao1990spin}%
  \BibitemOpen
  \bibfield  {author} {\bibinfo {author} {\bibfnamefont {X.}~\bibnamefont
  {Hao}}, \bibinfo {author} {\bibfnamefont {J.}~\bibnamefont {Moodera}},\ and\
  \bibinfo {author} {\bibfnamefont {R.}~\bibnamefont {Meservey}},\ }\bibfield
  {title} {\bibinfo {title} {Spin-filter effect of ferromagnetic europium
  sulfide tunnel barriers},\ }\href@noop {} {\bibfield  {journal} {\bibinfo
  {journal} {Physical Review B}\ }\textbf {\bibinfo {volume} {42}},\ \bibinfo
  {pages} {8235} (\bibinfo {year} {1990})}\BibitemShut {NoStop}%
\bibitem [{\citenamefont {Hao}\ \emph {et~al.}(1991)\citenamefont {Hao},
  \citenamefont {Moodera},\ and\ \citenamefont {Meservey}}]{hao1991thin}%
  \BibitemOpen
  \bibfield  {author} {\bibinfo {author} {\bibfnamefont {X.}~\bibnamefont
  {Hao}}, \bibinfo {author} {\bibfnamefont {J.}~\bibnamefont {Moodera}},\ and\
  \bibinfo {author} {\bibfnamefont {R.}~\bibnamefont {Meservey}},\ }\bibfield
  {title} {\bibinfo {title} {Thin-film superconductor in an exchange field},\
  }\href@noop {} {\bibfield  {journal} {\bibinfo  {journal} {Physical Review
  Letters}\ }\textbf {\bibinfo {volume} {67}},\ \bibinfo {pages} {1342}
  (\bibinfo {year} {1991})}\BibitemShut {NoStop}%
\bibitem [{\citenamefont {Strambini}\ \emph {et~al.}(2017)\citenamefont
  {Strambini}, \citenamefont {Golovach}, \citenamefont {De~Simoni},
  \citenamefont {Moodera}, \citenamefont {Bergeret},\ and\ \citenamefont
  {Giazotto}}]{PhysRevMaterials.1.054402}%
  \BibitemOpen
  \bibfield  {author} {\bibinfo {author} {\bibfnamefont {E.}~\bibnamefont
  {Strambini}}, \bibinfo {author} {\bibfnamefont {V.~N.}\ \bibnamefont
  {Golovach}}, \bibinfo {author} {\bibfnamefont {G.}~\bibnamefont {De~Simoni}},
  \bibinfo {author} {\bibfnamefont {J.~S.}\ \bibnamefont {Moodera}}, \bibinfo
  {author} {\bibfnamefont {F.~S.}\ \bibnamefont {Bergeret}},\ and\ \bibinfo
  {author} {\bibfnamefont {F.}~\bibnamefont {Giazotto}},\ }\bibfield  {title}
  {\bibinfo {title} {Revealing the magnetic proximity effect in {EuS/Al}
  bilayers through superconducting tunneling spectroscopy},\ }\href
  {https://doi.org/10.1103/PhysRevMaterials.1.054402} {\bibfield  {journal}
  {\bibinfo  {journal} {Phys. Rev. Mater.}\ }\textbf {\bibinfo {volume} {1}},\
  \bibinfo {pages} {054402} (\bibinfo {year} {2017})}\BibitemShut {NoStop}%
\bibitem [{\citenamefont {Gomez-Perez}\ \emph {et~al.}(2020)\citenamefont
  {Gomez-Perez}, \citenamefont {Zhang}, \citenamefont {Calavalle},
  \citenamefont {Ilyn}, \citenamefont {Gonz{\'a}lez-Orellana}, \citenamefont
  {Gobbi}, \citenamefont {Rogero}, \citenamefont {Chuvilin}, \citenamefont
  {Golovach}, \citenamefont {Hueso} \emph {et~al.}}]{gomez2020strong}%
  \BibitemOpen
  \bibfield  {author} {\bibinfo {author} {\bibfnamefont {J.~M.}\ \bibnamefont
  {Gomez-Perez}}, \bibinfo {author} {\bibfnamefont {X.-P.}\ \bibnamefont
  {Zhang}}, \bibinfo {author} {\bibfnamefont {F.}~\bibnamefont {Calavalle}},
  \bibinfo {author} {\bibfnamefont {M.}~\bibnamefont {Ilyn}}, \bibinfo {author}
  {\bibfnamefont {C.}~\bibnamefont {Gonz{\'a}lez-Orellana}}, \bibinfo {author}
  {\bibfnamefont {M.}~\bibnamefont {Gobbi}}, \bibinfo {author} {\bibfnamefont
  {C.}~\bibnamefont {Rogero}}, \bibinfo {author} {\bibfnamefont
  {A.}~\bibnamefont {Chuvilin}}, \bibinfo {author} {\bibfnamefont {V.~N.}\
  \bibnamefont {Golovach}}, \bibinfo {author} {\bibfnamefont {L.~E.}\
  \bibnamefont {Hueso}}, \emph {et~al.},\ }\bibfield  {title} {\bibinfo {title}
  {Strong interfacial exchange field in a heavy metal/ferromagnetic insulator
  system determined by spin {Hall} magnetoresistance},\ }\href@noop {}
  {\bibfield  {journal} {\bibinfo  {journal} {Nano Letters}\ }\textbf {\bibinfo
  {volume} {20}},\ \bibinfo {pages} {6815} (\bibinfo {year}
  {2020})}\BibitemShut {NoStop}%
\bibitem [{\citenamefont {Hauser}(1969)}]{hauser1969coupling}%
  \BibitemOpen
  \bibfield  {author} {\bibinfo {author} {\bibfnamefont {J.}~\bibnamefont
  {Hauser}},\ }\bibfield  {title} {\bibinfo {title} {Coupling between
  ferrimagnetic insulators through a superconducting layer},\ }\href@noop {}
  {\bibfield  {journal} {\bibinfo  {journal} {Physical Review Letters}\
  }\textbf {\bibinfo {volume} {23}},\ \bibinfo {pages} {374} (\bibinfo {year}
  {1969})}\BibitemShut {NoStop}%
\bibitem [{\citenamefont {Li}\ \emph {et~al.}(2013)\citenamefont {Li},
  \citenamefont {Roschewsky}, \citenamefont {Assaf}, \citenamefont {Eich},
  \citenamefont {Epstein-Martin}, \citenamefont {Heiman}, \citenamefont
  {M\"unzenberg},\ and\ \citenamefont {Moodera}}]{PhysRevLett.110.097001}%
  \BibitemOpen
  \bibfield  {author} {\bibinfo {author} {\bibfnamefont {B.}~\bibnamefont
  {Li}}, \bibinfo {author} {\bibfnamefont {N.}~\bibnamefont {Roschewsky}},
  \bibinfo {author} {\bibfnamefont {B.~A.}\ \bibnamefont {Assaf}}, \bibinfo
  {author} {\bibfnamefont {M.}~\bibnamefont {Eich}}, \bibinfo {author}
  {\bibfnamefont {M.}~\bibnamefont {Epstein-Martin}}, \bibinfo {author}
  {\bibfnamefont {D.}~\bibnamefont {Heiman}}, \bibinfo {author} {\bibfnamefont
  {M.}~\bibnamefont {M\"unzenberg}},\ and\ \bibinfo {author} {\bibfnamefont
  {J.~S.}\ \bibnamefont {Moodera}},\ }\bibfield  {title} {\bibinfo {title}
  {Superconducting spin switch with infinite magnetoresistance induced by an
  internal exchange field},\ }\href
  {https://doi.org/10.1103/PhysRevLett.110.097001} {\bibfield  {journal}
  {\bibinfo  {journal} {Phys. Rev. Lett.}\ }\textbf {\bibinfo {volume} {110}},\
  \bibinfo {pages} {097001} (\bibinfo {year} {2013})}\BibitemShut {NoStop}%
\bibitem [{\citenamefont {Zhu}\ \emph {et~al.}(2017)\citenamefont {Zhu},
  \citenamefont {Pal}, \citenamefont {Blamire},\ and\ \citenamefont
  {Barber}}]{zhu2017superconducting}%
  \BibitemOpen
  \bibfield  {author} {\bibinfo {author} {\bibfnamefont {Y.}~\bibnamefont
  {Zhu}}, \bibinfo {author} {\bibfnamefont {A.}~\bibnamefont {Pal}}, \bibinfo
  {author} {\bibfnamefont {M.~G.}\ \bibnamefont {Blamire}},\ and\ \bibinfo
  {author} {\bibfnamefont {Z.~H.}\ \bibnamefont {Barber}},\ }\bibfield  {title}
  {\bibinfo {title} {Superconducting exchange coupling between ferromagnets},\
  }\href@noop {} {\bibfield  {journal} {\bibinfo  {journal} {Nature Materials}\
  }\textbf {\bibinfo {volume} {16}},\ \bibinfo {pages} {195} (\bibinfo {year}
  {2017})}\BibitemShut {NoStop}%
\bibitem [{\citenamefont {Golubov}\ and\ \citenamefont
  {Kupriyanov}(2017)}]{golubov2017superconductivity43299613}%
  \BibitemOpen
  \bibfield  {author} {\bibinfo {author} {\bibfnamefont {A.~A.}\ \bibnamefont
  {Golubov}}\ and\ \bibinfo {author} {\bibfnamefont {M.~Yu.}\ \bibnamefont
  {Kupriyanov}},\ }\bibfield  {title} {\bibinfo {title} {Superconductivity:
  Controlling magnetism},\ }\href@noop {} {\bibfield  {journal} {\bibinfo
  {journal} {Nature Materials}\ }\textbf {\bibinfo {volume} {16}},\ \bibinfo
  {pages} {156} (\bibinfo {year} {2017})}\BibitemShut {NoStop}%
\bibitem [{\citenamefont {Matsuki}\ \emph {et~al.}(2025)\citenamefont
  {Matsuki}, \citenamefont {Hijano}, \citenamefont {Mazur}, \citenamefont
  {Ili{\'c}}, \citenamefont {Wang}, \citenamefont {Alekhina}, \citenamefont
  {Ohnishi}, \citenamefont {Komori}, \citenamefont {Li}, \citenamefont
  {Stelmashenko}, \citenamefont {Banerjee}, \citenamefont {Cohen},
  \citenamefont {McComb}, \citenamefont {Bergeret}, \citenamefont {Yang},\ and\
  \citenamefont {Robinson}}]{NatCommun2025}%
  \BibitemOpen
  \bibfield  {author} {\bibinfo {author} {\bibfnamefont {H.}~\bibnamefont
  {Matsuki}}, \bibinfo {author} {\bibfnamefont {A.}~\bibnamefont {Hijano}},
  \bibinfo {author} {\bibfnamefont {G.~P.}\ \bibnamefont {Mazur}}, \bibinfo
  {author} {\bibfnamefont {S.}~\bibnamefont {Ili{\'c}}}, \bibinfo {author}
  {\bibfnamefont {B.}~\bibnamefont {Wang}}, \bibinfo {author} {\bibfnamefont
  {I.}~\bibnamefont {Alekhina}}, \bibinfo {author} {\bibfnamefont
  {K.}~\bibnamefont {Ohnishi}}, \bibinfo {author} {\bibfnamefont
  {S.}~\bibnamefont {Komori}}, \bibinfo {author} {\bibfnamefont
  {Y.}~\bibnamefont {Li}}, \bibinfo {author} {\bibfnamefont {N.}~\bibnamefont
  {Stelmashenko}}, \bibinfo {author} {\bibfnamefont {N.}~\bibnamefont
  {Banerjee}}, \bibinfo {author} {\bibfnamefont {L.~F.}\ \bibnamefont {Cohen}},
  \bibinfo {author} {\bibfnamefont {D.~W.}\ \bibnamefont {McComb}}, \bibinfo
  {author} {\bibfnamefont {F.~S.}\ \bibnamefont {Bergeret}}, \bibinfo {author}
  {\bibfnamefont {G.}~\bibnamefont {Yang}},\ and\ \bibinfo {author}
  {\bibfnamefont {J.~W.~A.}\ \bibnamefont {Robinson}},\ }\bibfield  {title}
  {\bibinfo {title} {Realisation of de {Gennes'} absolute superconducting
  switch with a heavy metal interface},\ }\href
  {https://doi.org/10.1038/s41467-025-61267-2} {\bibfield  {journal} {\bibinfo
  {journal} {Nat. Commun.}\ }\textbf {\bibinfo {volume} {16}},\ \bibinfo
  {pages} {5674} (\bibinfo {year} {2025})}\BibitemShut {NoStop}%
\bibitem{SupplementalMaterial}%
  See the Supplemental Material included after the main text for the matrix Andreev-amplitude formulation, qualitative expectations for antiparallel alignment, numerical details, and additional figures.
\bibitem [{\citenamefont {Golubov}\ \emph {et~al.}(2002)\citenamefont
  {Golubov}, \citenamefont {Kupriyanov},\ and\ \citenamefont
  {Fominov}}]{Fominov2002}%
  \BibitemOpen
  \bibfield  {author} {\bibinfo {author} {\bibfnamefont {A.~A.}\ \bibnamefont
  {Golubov}}, \bibinfo {author} {\bibfnamefont {M.~Yu.}\ \bibnamefont
  {Kupriyanov}},\ and\ \bibinfo {author} {\bibfnamefont {Y.~V.}\ \bibnamefont
  {Fominov}},\ }\bibfield  {title} {\bibinfo {title} {Critical current in
  {SFIFS} junctions},\ }\href@noop {} {\bibfield  {journal} {\bibinfo
  {journal} {Pis'ma Zh. Eksp. Teor. Fiz.}\ }\textbf {\bibinfo {volume} {75}},\
  \bibinfo {pages} {223} (\bibinfo {year} {2002})},\ \bibinfo {note} {[JETP
  Letters 75, 190 (2002)]}\BibitemShut {NoStop}%
\bibitem [{\citenamefont {Dorokhov}(1984)}]{dorokhov1984}%
  \BibitemOpen
  \bibfield  {author} {\bibinfo {author} {\bibfnamefont {O.~N.}\ \bibnamefont
  {Dorokhov}},\ }\bibfield  {title} {\bibinfo {title} {On the coexistence of
  localized and extended electronic states in the metallic phase},\ }\href@noop
  {} {\bibfield  {journal} {\bibinfo  {journal} {Solid State Communications}\
  }\textbf {\bibinfo {volume} {51}},\ \bibinfo {pages} {381} (\bibinfo {year}
  {1984})}\BibitemShut {NoStop}%
\bibitem [{\citenamefont {Nazarov}(1994)}]{Nazarov}%
  \BibitemOpen
  \bibfield  {author} {\bibinfo {author} {\bibfnamefont {Y.~V.}\ \bibnamefont
  {Nazarov}},\ }\bibfield  {title} {\bibinfo {title} {Limits of universality in
  disordered conductors},\ }\href {https://doi.org/10.1103/PhysRevLett.73.134}
  {\bibfield  {journal} {\bibinfo  {journal} {Phys. Rev. Lett.}\ }\textbf
  {\bibinfo {volume} {73}},\ \bibinfo {pages} {134} (\bibinfo {year}
  {1994})}\BibitemShut {NoStop}%
\bibitem [{\citenamefont {Bardas}\ and\ \citenamefont
  {Averin}(1997)}]{Bardas1997}%
  \BibitemOpen
  \bibfield  {author} {\bibinfo {author} {\bibfnamefont {A.}~\bibnamefont
  {Bardas}}\ and\ \bibinfo {author} {\bibfnamefont {D.~V.}\ \bibnamefont
  {Averin}},\ }\bibfield  {title} {\bibinfo {title} {Electron transport in
  mesoscopic disordered superconductor--normal-metal--superconductor
  junctions},\ }\href {https://doi.org/10.1103/PhysRevB.56.R8518} {\bibfield
  {journal} {\bibinfo  {journal} {Physical Review B}\ }\textbf {\bibinfo
  {volume} {56}},\ \bibinfo {pages} {R8518} (\bibinfo {year}
  {1997})}\BibitemShut {NoStop}%
\bibitem [{\citenamefont {Beenakker}(1997)}]{Beenakker1997}%
  \BibitemOpen
  \bibfield  {author} {\bibinfo {author} {\bibfnamefont {C.~W.~J.}\
  \bibnamefont {Beenakker}},\ }\bibfield  {title} {\bibinfo {title}
  {Random-matrix theory of quantum transport},\ }\href@noop {} {\bibfield
  {journal} {\bibinfo  {journal} {Rev. Mod. Phys.}\ }\textbf {\bibinfo {volume}
  {69}},\ \bibinfo {pages} {731} (\bibinfo {year} {1997})}\BibitemShut
  {NoStop}%
\bibitem [{\citenamefont {Usadel}(1970)}]{Usadel1970}%
  \BibitemOpen
  \bibfield  {author} {\bibinfo {author} {\bibfnamefont {K.~D.}\ \bibnamefont
  {Usadel}},\ }\bibfield  {title} {\bibinfo {title} {Generalized diffusion
  equation for superconducting alloys},\ }\href
  {https://doi.org/10.1103/PhysRevLett.25.507} {\bibfield  {journal} {\bibinfo
  {journal} {Physical Review Letters}\ }\textbf {\bibinfo {volume} {25}},\
  \bibinfo {pages} {507} (\bibinfo {year} {1970})}\BibitemShut {NoStop}%
\bibitem [{\citenamefont {Kuprianov}\ and\ \citenamefont
  {Lukichev}(1988)}]{KupriyanovLukichev1988}%
  \BibitemOpen
  \bibfield  {author} {\bibinfo {author} {\bibfnamefont {M.~{\relax Yu}.}\
  \bibnamefont {Kuprianov}}\ and\ \bibinfo {author} {\bibfnamefont {V.~F.}\
  \bibnamefont {Lukichev}},\ }\bibfield  {title} {\bibinfo {title} {Influence
  of boundary transparency on the critical current of ``dirty'' {SS'S}
  structures},\ }\href@noop {} {\bibfield  {journal} {\bibinfo  {journal}
  {Zh. Eksp. Teor. Fiz.}\ }\textbf {\bibinfo {volume} {94}},\ \bibinfo {pages}
  {139} (\bibinfo {year} {1988})}\ \bibinfo {note} {[Sov. Phys. JETP
  \textbf {67}, 1163 (1988)]}\BibitemShut {NoStop}%
\end{thebibliography}
\end{document}

\onecolumngrid
\setcounter{suppsection}{0}
\setcounter{suppsubsection}{0}
\renewcommand{\theHequation}{supp.\arabic{equation}}
\renewcommand{\theHfigure}{supp.\arabic{figure}}
\begin{center}
{\bfseries\large Supplemental Material for\\[0.4em]
Multiple Andreev Reflection Spectroscopy of Spin-Resolved Proximity States\par}
\vspace{0.5em}
A.~A.~Golubov and M.~Yu.~Kupriyanov\\[0.5em]
{\bfseries Matrix Andreev Reflection Amplitudes and the Averin--Bardas MAR Scheme in SFcFS Junctions\par}
\end{center}
\vspace{0.5em}

\renewcommand{\theequation}{S\arabic{equation}}
\renewcommand{\thefigure}{S\arabic{figure}}
\setcounter{equation}{0}
\setcounter{figure}{0}

\section*{Overview}

The present SFcFS model combines three theoretical ingredients.

(i) The constriction is assumed to be short and ballistic and is described by the \begingroup Averin--Bardas (AB) theory of coherent multiple Andreev reflections (MAR)\endgroup{}~\cite{supp:AverinBardas1995}. Its normal-state properties are characterized by the transmission coefficient $D$.

(ii) The superconducting electrodes are equilibrium spectral reservoirs. They are assumed to be diffusive and are described by quasiclassical Green functions obtained from the Usadel equations.

(iii) The quantity connecting these two descriptions is the local Andreev reflection amplitude $a(E)$. This amplitude enters the Averin--Bardas recurrence relations and can be expressed through the quasiclassical Green functions of the diffusive electrodes.

Conceptually, the construction may be summarized as

\[
\text{Usadel electrodes}
\rightarrow
(G_\sigma,F_\sigma)
\rightarrow
a_\sigma(E)
\rightarrow
\text{AB recurrence}
\rightarrow
I(V).
\]
Schopohl's Riccati formulation provides the microscopic interpretation of the same quantity $a(E)$ in terms of local quasiparticle amplitudes.

\SuppSection{From the Averin--Bardas MAR scheme to matrix Andreev amplitudes} The purpose of this Supplementary Material is to show how the conventional Averin--Bardas description of coherent multiple Andreev reflections (MAR) is generalized from a scalar Andreev amplitude \(a(E)\) to side-dependent spin matrices \(\hat a_L(E)\) and \(\hat a_R(E)\). This formulation naturally includes arbitrary spin-dependent superconducting electrodes. The parallel SFcFS geometry studied in the main text is then obtained as the simplest diagnostic limit of this more general matrix framework. The Averin--Bardas formulation of coherent MAR in a short single-channel constriction requires two ingredients. The first is the normal-state scattering matrix of the constriction, \begin{equation} S_N = \begin{pmatrix} r & t \\ t & r \end{pmatrix}, \qquad |t|^2=D,\qquad |r|^2=1-D , \label{eq:SN_supp} \end{equation} where \(D\) is the normal-state transparency.
\begingroup
The phases also obey $rt^*+tr^*=0$, as required by unitarity.
The real scattering matrix used in Appendix B of the main text is an
equivalent channel-phase convention.\endgroup{} The second ingredient is the Andreev reflection amplitude of the superconducting banks. In a voltage-biased junction the MAR ladder contains the shifted energies \begin{equation} E_n = E+n eV , \label{eq:En_supp} \end{equation} and the recurrence relates quasiparticle amplitudes at neighboring rungs of this ladder through the normal scattering amplitudes \(r,t\) and the Andreev amplitudes evaluated at \(E_n\). For ordinary identical BCS electrodes the spectral input is a scalar amplitude \(a_{\rm BCS}(E_n)\). More generally, even for spin-independent normal scattering, the superconducting banks may have different spin-dependent spectra. The natural generalization is therefore \begin{equation} a_{\rm BCS}(E_n) \quad \longrightarrow \quad \hat a_{\alpha,n} \equiv \hat a_\alpha(E_n), \qquad \alpha=L,R , \label{eq:scalar_to_matrix_supp} \end{equation} where \(\hat a_L\) and \(\hat a_R\) are \(2\times2\) matrices in spin space attached to the left and right electrodes.

In the quasiclassical formulation used below, these amplitudes are
expressed through the dimensionless normal and anomalous retarded
quasiclassical Green functions $\hat G_\alpha^R(E)$ and
$\hat F_\alpha^R(E)$, respectively, evaluated at the constriction side
of electrode $\alpha=L,R$:

\begin{equation} \hat a_{\alpha,n} = i\hat F^R_\alpha(E_n) \left[ \hat 1+\hat G^R_\alpha(E_n) \right]^{-1}. \label{eq:a_matrix_side_supp} \end{equation} Equation~\eqref{eq:a_matrix_side_supp} is the point at which the spin-dependent superconducting spectra enter the coherent MAR problem. The derivation of this relation from the local Andreev equations and from the quasiclassical Green functions is given in Sec.~II. In the most general magnetic configuration the matrices \(\hat a_{L,n}\) and \(\hat a_{R,n}\) need not commute with each other or with the spin structure of the normal-state scattering matrix. The MAR recurrence is then genuinely matrix-valued in spin space. A full noncollinear recurrence is not required for the present calculation, but Eq.~\eqref{eq:a_matrix_side_supp} makes clear how such a generalization is organized: the scalar products of Andreev amplitudes in the original recurrence are replaced by ordered products of the corresponding spin matrices. For collinear magnetizations, the spin quantization axis is common to both electrodes and the matrices become diagonal, \begin{equation} \hat a_{\alpha,n} = \begin{pmatrix} a_{\alpha,n,+} & 0 \\ 0 & a_{\alpha,n,-} \end{pmatrix}, \qquad \alpha=L,R . \label{eq:diag_a_supp} \end{equation} The side-resolved scalar amplitudes are \begin{equation} a_{\alpha,n,\sigma} = \frac{ iF^R_{\alpha,\sigma}(E_n) }{ 1+G^R_{\alpha,\sigma}(E_n) }, \qquad \sigma=\pm . \label{eq:side_resolved_a_supp} \end{equation} A quasiparticle undergoing MAR alternately experiences Andreev reflection at the left and right banks. Therefore, even in a collinear junction, an asymmetric SFcFS structure is generally described by side-dependent products such as \begin{equation} a_{L,n,\sigma}a_{R,n+1,\sigma}, \qquad a_{R,n,\sigma}a_{L,n+1,\sigma}. \label{eq:side_products_supp} \end{equation} Thus a collinear asymmetric junction cannot, in general, be reduced to a single scalar Andreev amplitude. Antiparallel SFcFS spin valves, for which \(H_L=-H_R\), belong to this side-resolved class. 

\SuppSubsection{Qualitative expectation for antiparallel SFcFS spin valves}
\label{sec:ap_qualitative_supp}

The side-resolved formulation also clarifies what should be expected
qualitatively in an antiparallel SFcFS spin valve. For
\(H_L=H\) and \(H_R=-H\), the left and right electrodes are described
by different spin-resolved amplitudes,
\begin{equation}
    a_{L,n,\sigma}=a_\sigma(E_n,H),
    \qquad
    a_{R,n,\sigma}=a_\sigma(E_n,-H).
    \label{eq:AP_side_amplitudes_supp}
\end{equation}
Equivalently, with the spin convention used in the main text,
reversing the exchange field interchanges the two spin-resolved
spectral functions. The MAR recurrence is therefore side-resolved even
though the magnetizations are collinear.

A simple spectral-interval argument gives the expected scale of the
effect. In the parallel case studied in the main text, the shifted MAR
component is associated with the local reconstructed interval
\(\Delta+E_p\), where \(E_p=|E_{\rm peak}|\). Thus
\begingroup
\begin{equation}
    eV_n^{(p)}
    \simeq
    \frac{\Delta+E_p}{n},
    \qquad
    e\,\Delta V_n^{\rm P}
    \simeq
    \frac{\Delta-E_p}{n}.
    \label{eq:P_scale_supp}
\end{equation}
\endgroup
In the antiparallel case, odd-order trajectories connect opposite
electrodes. Since the two electrodes contain mirror-reconstructed
spectral singularities, such trajectories can probe the interval
between the two shifted singularities, approximately \(2E_p\), in
addition to the conventional interval \(2\Delta\). Therefore one
expects an additional odd-order scale
\begingroup
\begin{equation}
    eV_{n,{\rm AP}}^{(p)}
    \simeq
    \frac{2E_p}{n},
    \qquad
    e\,\Delta V_n^{\rm AP,odd}
    \simeq
    \frac{2(\Delta-E_p)}{n}.
    \label{eq:AP_odd_scale_supp}
\end{equation}
\endgroup
At the level of this kinematic estimate, the odd-order antiparallel
splitting is therefore expected to be larger than the corresponding
parallel splitting by approximately a factor of two.

Even-order trajectories return to the same electrode and are not
expected to acquire this cross-electrode \(2E_p\) scale. Nevertheless,
unlike in a rigid Zeeman-shifted BCS model, they can still probe the
local reconstructed interval \(\Delta+E_p\). Thus the even-order
antiparallel anomalies should remain modified on the same scale as in
the parallel case,
\begingroup
\begin{equation}
    e\,\Delta V_n^{\rm AP,even}
    \simeq
    \frac{\Delta-E_p}{n},
    \label{eq:AP_even_scale_supp}
\end{equation}
\endgroup
although their spectral weight and detailed line shape require the
full side-resolved recurrence.

Equations~\eqref{eq:AP_odd_scale_supp} and
\eqref{eq:AP_even_scale_supp} should be understood as qualitative
spectral-interval estimates, not as a substitute for a full numerical
calculation. The actual peak positions and intensities depend on the
ordered products of side-resolved amplitudes in
Eq.~\eqref{eq:side_products_supp}. They nevertheless indicate that an
antiparallel SFcFS spin valve should exhibit a richer MAR spectrum
than the parallel junction: odd-order anomalies are expected to show a
larger cross-electrode splitting, while even-order anomalies should
still retain the proximity-reconstruction scale.

\SuppSubsection{Parallel SFcFS spin valves}
\label{sec:ap_parallel_supp}

The main text focuses on the symmetric parallel case,
\begin{equation}
    H_L=H_R=H, \qquad
    \hat G^R_L=\hat G^R_R, \qquad
    \hat F^R_L=\hat F^R_R .
    \label{eq:parallel_condition_supp}
\end{equation}
Then
\begin{equation}
    a_{L,n,\sigma}=a_{R,n,\sigma}
    \equiv a_\sigma(E_n,H),
    \label{eq:parallel_a_supp}
\end{equation}
and the generalized recurrence collapses exactly to two independent
copies of the scalar Averin--Bardas recurrence,
\begin{equation}
    {\cal R}_{\rm AB}[a_+] \oplus {\cal R}_{\rm AB}[a_-].
    \label{eq:parallel_recurrence_supp}
\end{equation}
Here \({\cal R}_{\rm AB}[a_\sigma]\) denotes the scalar
Averin--Bardas recurrence evaluated with the energy-dependent Andreev
amplitude \(a_\sigma(E_n)\).

The parallel configuration is therefore the cleanest limit in which the
general matrix framework reduces to two scalar recurrences while
retaining the non-BCS spin-resolved spectral input of the SF electrodes.
It is also the most direct diagnostic geometry: for a rigid
Zeeman-shifted BCS spectrum with parallel spin splittings the conventional
MAR thresholds remain unchanged, whereas the proximity-reconstructed SF
spectrum considered here redistributes and splits the subharmonic-gap
anomalies. In the rigid-Zeeman case the spin-dependent shift is common
to the two electrodes in each sector and can be absorbed into the MAR
energy variable; it therefore does not introduce an additional spectral
interval in the parallel configuration.

For the symmetric parallel SFcFS junction, the two spin-resolved spectral functions are related by particle--hole symmetry. With the conventions used in the main text, \begin{equation} G^R_+(E,H) = \left[ G^R_-(-E,H) \right]^*, \qquad F^R_+(E,H) = \left[ F^R_-(-E,H) \right]^* . \label{eq:spectral_symmetry_supp} \end{equation} Consequently, the corresponding Andreev amplitudes obey \begin{equation} a_+(E,H) = - \left[ a_-(-E,H) \right]^* , \label{eq:a_symmetry_supp} \end{equation} and the spin-resolved densities of states satisfy \begin{equation} N_+(E,H)=N_-(-E,H). \label{eq:DOS_symmetry_supp} \end{equation} Thus, if the \(+\) sector contains a shifted negative-energy singularity at \(E=-E_p\), the \(-\) sector contains the mirror singularity at \(E=+E_p\). These relations do not imply that the sector-resolved MAR currents are equal. The two MAR ladders sample different energy-dependent spectral functions, and in general \begin{equation} I_+(V,H)\neq I_-(V,H). \label{eq:sector_currents_unequal_supp} \end{equation} Moreover, the sector-resolved quantities are not separately measurable charge currents. Bias reversal interchanges the two Nambu-spin sectors, \begin{equation} I_+(-V,H)=-I_-(V,H), \qquad I_-(-V,H)=-I_+(V,H). \label{eq:bias_sector_symmetry_supp} \end{equation} Therefore the physical current, \begin{equation} I(V,H) = \frac{1}{2} \left[ I_+(V,H)+I_-(V,H) \right], \label{eq:physical_current_supp} \end{equation} is antisymmetric, \begin{equation} I(-V,H)=-I(V,H). \label{eq:total_current_antisymmetry_supp} \end{equation} This provides a useful check on the sector-resolved MAR calculation.

\SuppSubsection{Explicit scalar Averin--Bardas recurrence used in each sector}
\label{sec:explicit_AB_supp}

For the symmetric parallel SFcFS junction the general matrix recurrence
reduces to two independent scalar Averin--Bardas recurrences, 
\({\cal R}_{\rm AB}[a_+]\) and \({\cal R}_{\rm AB}[a_-]\).
We now write the scalar
recurrence explicitly. The notation follows the original
formulation~\cite{supp:AverinBardas1995}, with the only change that the BCS
Andreev amplitude is replaced by the spin-resolved proximity amplitude,
\begin{equation}
    a_m
    \longrightarrow
    a_{\sigma,m}
    \equiv
    a_\sigma(E+meV,H),
    \qquad
    \sigma=\pm .
    \label{eq:a_m_sigma_supp}
\end{equation}
For a channel of transparency \(D\), we denote \(R=1-D\). The source
amplitude for a quasiparticle incident from the left electrode is
\begin{equation}
    J_\sigma(E)
    =
    \left[
        1-\left|a_{\sigma,0}\right|^2
    \right]^{1/2}.
    \label{eq:J_sigma_supp}
\end{equation}

The amplitudes \(A_{n,\sigma}\) and \(B_{n,\sigma}\) of the MAR ladder
obey the coupled recurrence relations
\begin{align}
&
\frac{
    D\,a_{\sigma,2n+2}a_{\sigma,2n+1}
}{
    1-a_{\sigma,2n+1}^{\,2}
}
B_{n+1,\sigma}
-
\left[
D\left(
\frac{
    a_{\sigma,2n+1}^{\,2}
}{
    1-a_{\sigma,2n+1}^{\,2}
}
+
\frac{
    a_{\sigma,2n}^{\,2}
}{
    1-a_{\sigma,2n-1}^{\,2}
}
\right)
+
1-a_{\sigma,2n}^{\,2}
\right]
B_{n,\sigma}
\nonumber\\
&\hspace{3.0cm}
+
\frac{
    D\,a_{\sigma,2n}a_{\sigma,2n-1}
}{
    1-a_{\sigma,2n-1}^{\,2}
}
B_{n-1,\sigma}
=
-\sqrt{R}\,J_\sigma(E)\,\delta_{n0},
\label{eq:AB_B_recurrence_supp}
\end{align}
and
\begin{equation}
    A_{n+1,\sigma}
    -
    a_{\sigma,2n+1}a_{\sigma,2n}A_{n,\sigma}
    =
    \sqrt{R}
    \left(
        B_{n+1,\sigma}a_{\sigma,2n+2}
        -
        B_{n,\sigma}a_{\sigma,2n+1}
    \right)
    +
    J_\sigma(E)a_{\sigma,1}\delta_{n0}.
    \label{eq:AB_A_recurrence_supp}
\end{equation}
Equations~\eqref{eq:AB_B_recurrence_supp} and
\eqref{eq:AB_A_recurrence_supp} are solved on a finite ladder
\(-N\le n\le N\), with \(N\) increased until the current is converged
for the voltage under consideration.

\begingroup
To specify the current completely, define the kernel in the
source-containing convention of
Eqs.~\eqref{eq:AB_B_recurrence_supp} and
\eqref{eq:AB_A_recurrence_supp}:
\[
\begin{aligned}
\mathcal K_\sigma(E,V;D)
 &=2J_\sigma\operatorname{Re}
       [a_{\sigma,0}A_{0,\sigma}]\\
 &\quad+\sum_n(1+|a_{\sigma,2n}|^2)
 (|A_{n,\sigma}|^2-|B_{n,\sigma}|^2).
\end{aligned}
\]
Here $A$ and $B$ already contain $J_\sigma$; multiplying this
expression by an additional $1-|a_{\sigma,0}|^2$ would count the
source twice. With $R_Q=h/(2e^2)$ and the spin-degenerate channel
resistance $R_N(D)=R_Q/D$, the sector-current convention of the main
text is
\begin{equation}
 I_\sigma(V,T;D)=\frac{1}{eR_Q}
 \left[DeV-\lim_{E_c\to\infty}\int_{-E_c}^{E_c}dE\,
 \tanh\!\left(\frac{E}{2T}\right)\mathcal K_\sigma(E,V;D)\right].
 \label{eq:I_sigma_AB_supp}
\end{equation}
Appendix B of the main text derives this symmetric-cutoff expression,
explains the explicit normal-current term $DeV$, and gives its
equivalent unit-source form. The two formal sector quantities carry the
spin-degenerate prefactor; the experimentally relevant charge current
is therefore their average, not their unweighted sum. In particular,
the normal limit gives $I=V/R_N(D)$.\endgroup{}
\begin{equation}
    I(V,T;D)
    =
    \frac{1}{2}
    \left[
        I_+(V,T;D)+I_-(V,T;D)
    \right].
    \label{eq:I_total_AB_supp}
\end{equation}
Thus the formal structure of the Averin--Bardas recurrence is unchanged;
all information about the spin-resolved proximity spectrum enters
through the replacement \(a_m\to a_{\sigma,m}\).

\SuppSection{Matrix Andreev amplitude from quasiclassical Green functions} \label{sec:matrix_amplitude} In this section we justify the matrix Andreev amplitude used in Sec.~I. The derivation is local: it refers to a single superconducting electrode and does not depend on the details of the constriction or on the particular SF model used later. The result is the spin-matrix generalization of the relation between the local Andreev reflection amplitude and the retarded quasiclassical Green functions~\begingroup\cite{supp:GolubovKupriyanov1995,supp:Schopohl1998,supp:Zaitsev1998}\endgroup{}. \SuppSubsection{Local coherence amplitude} In the quasiclassical approximation the rapidly oscillating Fermi-scale phase is separated from the quasiparticle wave function. Along a trajectory normal to the constriction, the remaining slowly varying electron-like and hole-like amplitudes obey local Andreev equations. For a spin-independent electrode one may write \begin{equation} -i\hbar v_F \partial_x \begin{pmatrix} u \\ v \end{pmatrix} = \begin{pmatrix} E+i0 & -\Delta \\ \Delta^* & -E-i0 \end{pmatrix} \begin{pmatrix} u \\ v \end{pmatrix}. \label{eq:local_andreev_scalar} \end{equation} The local Andreev amplitude is the ratio between the outgoing hole-like and incoming electron-like components. For a spin-dependent electrode, \(u\) and \(v\) are spinors. The local Andreev equation becomes \begin{equation} -i\hbar v_F \partial_x \begin{pmatrix} u \\ v \end{pmatrix} = \begin{pmatrix} E+i0-\hat h & -\hat \Delta \\ \hat \Delta^\dagger & -E-i0-\hat h^T \end{pmatrix} \begin{pmatrix} u \\ v \end{pmatrix}, \label{eq:local_andreev_matrix} \end{equation} where \(\hat h={\bf H}\cdot\hat{\boldsymbol\sigma}\) and \(\hat\Delta=\Delta i\hat\sigma_y\) for singlet pairing. \begingroup
A hat denotes a $2\times2$ spin matrix, a check denotes a matrix in
Nambu and spin space, and the superscript $\mathsf T$ (written $T$ in
the equations) denotes ordinary transposition, not time reversal.
\endgroup{} \begingroup
From this point we express holes in the time-reversed basis
$\bar v=i\hat\sigma_yv=(v_\downarrow,-v_\uparrow)^T$, so that the
singlet spin matrix is absorbed into the hole coordinates. A local
outgoing retarded solution defines
\begin{equation}
 \bar v=\hat a(E)u.
 \label{eq:def_matrix_a}
\end{equation}
Thus $\hat a$ maps an electron spinor to this hole spinor. In the
original physical-spin coordinates the same conversion is
$v=-i\hat\sigma_y\hat a u$; it is not a spin-conserving electron-to-hole
conversion. The diagonal collinear amplitudes below always refer to
the time-reversed hole basis.\endgroup{} In Riccati language it is the local coherence amplitude associated with the retarded branch of the quasiclassical solution. \SuppSubsection{Relation to the local Gor'kov amplitudes} The same object appears in the construction of the local Gor'kov solution near the constriction. In the spin-degenerate case the slowly varying parts of the electron-like and hole-like components may be written as \(g(E)\) and \(f(E)\), so that the Andreev amplitude is \begin{equation} a(E)=\frac{f(E)}{g(E)} . \label{eq:scalar_fg_ratio} \end{equation} For a spin-dependent electrode the slowly varying amplitudes become matrices in spin space, \(\hat g(E)\) and \(\hat f(E)\). For an arbitrary incident spinor \(c\), \begingroup\begin{equation} u=\hat g\,c, \qquad \bar v=\hat f\,c . \label{eq:matrix_gf_spinor} \end{equation}\endgroup{} 
Since \(c\) is arbitrary, comparison with Eq.~\eqref{eq:def_matrix_a} gives \begin{equation} \hat a(E)=\hat f(E)\hat g^{-1}(E). \label{eq:matrix_fg_ratio} \end{equation} 
The above equation is the direct spin-matrix analogue of the $f/g$ scalar ratio  ~\cite{supp:GolubovKupriyanov1995,supp:Schopohl1998,supp:Zaitsev1998}. 
Here $\hat g$ and $\hat f$ are wave-amplitude matrices, not the
normalized quasiclassical Green functions $\hat G^R$ and $\hat F^R$.
In the projector basis introduced in the following subsection, they
can be chosen as
\[
 \hat g=\frac{\hat1+\hat G^R}{2},\qquad
 \hat f=\frac{i\hat F^R}{2}.
\]
A common change of solution basis multiplies both matrices from the
right and leaves $\hat f\hat g^{-1}$ unchanged. 
Section~II\,C derives an equivalent expression for $\hat a$ in terms
of the retarded quasiclassical Green functions; no additional
normalization of $\hat a$ is involved.

\SuppSubsection{Expression through retarded quasiclassical Green functions} 
\begingroup
We now express this amplitude through the normalized retarded
quasiclassical propagator. In the electron/time-reversed-hole
coordinates and the anomalous convention $F^R=-i\Delta/Q$, write
\begin{equation}
 \check g^R=
 \begin{pmatrix}
 \hat G^R&-i\hat{\widetilde F}^{\,R}\\
 i\hat F^R&-\hat{\widetilde G}^{\,R}
 \end{pmatrix}.
 \label{eq:quasiclassical_matrix}
\end{equation}
Each block is a $2\times2$ spin matrix. The explicit factors $i$ and
the lower normal-block sign specify the convention: $i\hat F^R$ is
the electron-to-hole block. They must not be dropped when associating
the scalar $F^R$ with a raw Nambu-matrix element.\endgroup{} The normalization condition is \begin{equation} \left(\check g^R\right)^2=\check 1 . \label{eq:normalization_matrix} \end{equation} \begingroup
The normalized physical retarded solution defines the algebraic
projector
\[
 \check P_+=\frac{\check1+\check g^R}{2},\qquad
 \check P_+^2=\check P_+.
\]
It selects the electron-like retarded solution subspace; the label
$+$ here is a projector eigenvalue, not the spin-sector index $\sigma$.
The retarded transport equations and the electrode boundary conditions
select the physical $\check g^R$: normalization alone is not a boundary
condition. Acting on a column with electron coefficient $\chi$ gives
\[
 \begin{pmatrix}u\\\bar v\end{pmatrix}
 =\check P_+\begin{pmatrix}\chi\\0\end{pmatrix}
 =\frac12\begin{pmatrix}(\hat1+\hat G^R)\chi\\i\hat F^R\chi\end{pmatrix}.
\]
Where $\hat1+\hat G^R$ is invertible, this column parametrizes the
subspace and directly yields $\bar v=i\hat F^R(\hat1+\hat G^R)^{-1}u$.
Singular energies are understood by retarded continuation. This is
the component derivation used in Sec.~II of the main text. It agrees
with the coherence-function construction below, without requiring
commuting spin matrices. The identity is local and stationary; it
supplies electrode coherence, not a substitute for interface-scattering
boundary conditions. Its use with Usadel functions assumes the
isotropic diffusive limit.

A constructive way to enforce this normalization is to use the two
complementary solution subspaces $(u,\hat a u)^T$ and
$(\hat{\widetilde a}w,w)^T$. Set
\[
 \check Y=\begin{pmatrix}\hat1&\hat{\widetilde a}\\
                         \hat a&\hat1\end{pmatrix},\qquad
 \check g^R=\check Y\begin{pmatrix}\hat1&0\\0&-\hat1\end{pmatrix}
                   \check Y^{-1}.
\]
Where the inverse exists this gives $(\check g^R)^2=\check1$
identically, with singular energies understood by retarded limits.
Writing $\hat{\mathcal N}=(\hat1-\hat{\widetilde a}\hat a)^{-1}$
and $\hat{\widetilde{\mathcal N}}=(\hat1-\hat a\hat{\widetilde a})^{-1}$,
the upper normal and lower conversion blocks yield the following
relations. The complementary blocks are
$\hat{\widetilde G}^{\,R}=2\hat{\widetilde{\mathcal N}}-\hat1$ and
$\hat{\widetilde F}^{\,R}=-2i\hat{\widetilde a}\hat{\widetilde{\mathcal N}}$.
In the second solution subspace the hole spinor $w$ is specified,
and $\hat{\widetilde a}w$ is its electron component. Thus
$\hat a$ describes electron-to-hole coherence and
$\hat{\widetilde a}$ its hole-to-electron counterpart. The tilde
labels this companion function; it does not denote transposition.
Particle--hole symmetry relates the corresponding functions with
energy and trajectory reversal and the appropriate spin-basis
transformations, not by complex conjugation at the same energy.
The physical functions satisfy these symmetries and the retarded
electrode boundary conditions.\endgroup{}

\begingroup Thus the normal and anomalous conversion functions can be written as\endgroup{} \begin{equation} \hat G^R = \left(\hat 1-\hat{\widetilde a}\hat a\right)^{-1} \left(\hat 1+\hat{\widetilde a}\hat a\right), \label{eq:G_coherence_param} \end{equation} and \begin{equation} \hat F^R = -2i\, \hat a \left(\hat 1-\hat{\widetilde a}\hat a\right)^{-1}. \label{eq:F_coherence_param} \end{equation} This convention is chosen so that the scalar BCS limit reproduces the Averin--Bardas Andreev amplitude used in the main text. From Eq.~\eqref{eq:G_coherence_param}, \begin{equation} \hat 1+\hat G^R = 2\left(\hat 1-\hat{\widetilde a}\hat a\right)^{-1}. \label{eq:one_plus_G} \end{equation} Multiplying Eq.~\eqref{eq:F_coherence_param} by \((\hat 1+\hat G^R)^{-1}\) from the right gives \begin{equation} \hat F^R \left(\hat 1+\hat G^R\right)^{-1} = -i\hat a . \label{eq:F_times_inverse} \end{equation} Therefore, \begin{equation} \boxed{ \hat a(E) = i\hat F^R(E) \left[ \hat 1+\hat G^R(E) \right]^{-1} } . \label{eq:matrix_a_GF} \end{equation} This is the matrix Andreev amplitude entering the generalized MAR recurrence of Sec.~I. The order of the matrices in Eq.~\eqref{eq:matrix_a_GF} corresponds to the convention in which \(\hat a\) acts on the incident electron spinor from the left, as in Eq.~\eqref{eq:def_matrix_a}. In the collinear case all matrices are diagonal and the ordering becomes immaterial. \SuppSubsection{Scalar BCS limit} For a scalar BCS electrode, \begin{equation} G^R_{\rm BCS}(E) = \frac{E+i\Gamma} {\sqrt{(E+i\Gamma)^2-\Delta^2}}, \qquad F^R_{\rm BCS}(E) = -\frac{i\Delta} {\sqrt{(E+i\Gamma)^2-\Delta^2}} . \label{eq:BCS_GF_supp} \end{equation} Substitution into Eq.~\eqref{eq:matrix_a_GF} gives \begin{equation} a_{\rm BCS}(E) = \frac{ E+i\Gamma - \sqrt{(E+i\Gamma)^2-\Delta^2} }{ \Delta }, \label{eq:a_BCS_supp} \end{equation} which is the standard Andreev amplitude entering the \begingroup Blonder--Tinkham--Klapwijk (BTK)\endgroup{} and Averin--Bardas formulations. This check fixes the phase convention used above. \SuppSubsection{Collinear reduction} For a homogeneous collinear exchange field the spin quantization axis is fixed. In this basis the Gor'kov equations separate into two Nambu-spin sectors, \begingroup\begin{equation} \Psi_+ = \begin{pmatrix} u_\uparrow \\ v_\downarrow \end{pmatrix}, \qquad \Psi_- = \begin{pmatrix} u_\downarrow \\ -v_\uparrow \end{pmatrix}. \label{eq:nambu_spin_sectors_supp} \end{equation}\endgroup{} \begingroup
The minus sign is the time-reversed-hole basis convention, not a
change of the physical sector $(e_\downarrow,h_\uparrow)$. Collinear
exchange and spin-independent scattering cannot connect these two
sectors.\endgroup{} The quasiclassical Green functions are diagonal in this sector index, \begin{equation} \hat G^R = \begin{pmatrix} G^R_+ & 0 \\ 0 & G^R_- \end{pmatrix}, \qquad \hat F^R = \begin{pmatrix} F^R_+ & 0 \\ 0 & F^R_- \end{pmatrix}. \label{eq:diagonal_GF_supp} \end{equation} Equation~\eqref{eq:matrix_a_GF} then reduces to \begin{equation} a_\sigma(E) = \frac{ iF^R_\sigma(E) }{ 1+G^R_\sigma(E) }, \qquad \sigma=\pm . \label{eq:scalar_a_sigma_supp} \end{equation} The Andreev conversion \(e_\uparrow\leftrightarrow h_\downarrow\) takes place entirely within \(\Psi_+\), while \(e_\downarrow\leftrightarrow h_\uparrow\) takes place entirely within \(\Psi_-\). The index \(\sigma\) therefore labels a Nambu-spin sector, not a conserved electron spin. This is why the parallel SFcFS junction can be evaluated as two independent scalar MAR recurrences, even though each Andreev reflection reverses the physical spin of the quasiparticle.

\SuppSection{Thin-F SF electrode used in the main text}
\label{sec:thinF_supp}

\begingroup
The electrode Green functions used in the main text are the known
thin-F limit of an equilibrium SF bilayer~\cite{supp:Fominov2002}. Appendix A
of the main text gives the Usadel equation, SF boundary condition,
and thin-layer reduction explicitly. Here we retain the spectral
expressions and their conventions for use in the MAR recurrence. The F
layer is diffusive, with a homogeneous collinear exchange field and no
intrinsic pair potential. The S layer is a rigid bulk BCS reservoir;
spin relaxation, orbital depairing, and spin-dependent SF-interface
scattering are neglected. The small constriction is assumed not to
modify the equilibrium electrode spectrum.

\endgroup
We briefly recall the derivation following
Ref.~\cite{supp:Fominov2002}. In the F layer the Usadel functions obey
the normalization condition
\begin{equation}
    G_{\omega,\sigma}^2
    +
    F_{\omega,\sigma}F^*_{-\omega,\sigma}
    =
    1 ,
    \label{eq:usadel_norm_supp}
\end{equation}
and may be parameterized through the anomalous function
\(\Phi_{\omega,\sigma}\). The exchange field enters through the shifted
Matsubara frequency
\begin{equation}
    \widetilde \omega_\sigma
    =
    \omega+i\sigma H .
    \label{eq:omega_tilde_supp}
\end{equation}
\begingroup
For positive Matsubara frequency, the thin-layer reduction with the
spin-independent Kupriyanov--Lukichev SF boundary condition and a
reflecting outer boundary gives Eq.~(12) of
Ref.~\cite{supp:Fominov2002}, with the parent-superconductor phase factored
out:
\begin{equation}
    \Phi_{F,\sigma}(\omega)
    =
    \frac{\widetilde\omega_\sigma\Delta}
    {\displaystyle \omega+
    \frac{\gamma_B\widetilde\omega_\sigma}{\pi T_c}
    \sqrt{\omega^2+\Delta^2}},
    \qquad
    G_S=\frac{\omega}{\sqrt{\omega^2+\Delta^2}}.
    \label{eq:PhiF_thinF_supp}
\end{equation}
Here $\gamma_B=(R_{SF}/\rho_F\xi_F)(d_F/\xi_F)
=R_{SF}d_F/(\rho_F\xi_F^2)$ and
$\xi_F=\sqrt{\mathcal D_F/(2\pi T_c)}$, with $\hbar=k_B=1$.
The symbol $R_{SF}$ denotes the specific resistance of SF interface,
$\rho_F$ the F-layer resistivity, and $\mathcal D_F$ its diffusion
constant. The effective parameter $\gamma_B$ corresponds to
$\gamma_{BM}$ of Ref.~\cite{supp:Fominov2002}.
The thin-layer condition is
$d_F\ll\min(\xi_F,\sqrt{\mathcal D_F/(2|H|)})$ over the relevant
energy range. Small effective $\gamma_B$ alone does not fix the
microscopic interface transparency.

Writing the denominator of Eq.~\eqref{eq:PhiF_thinF_supp} as
\[
 W_\sigma(\omega)=\omega+
 \frac{\gamma_B\widetilde\omega_\sigma}{\pi T_c}
 \sqrt{\omega^2+\Delta^2},
\]
the continuation $\omega\mapsto-iz$, with $z=E+i0^+$, gives
\[
 W_\sigma(-iz)=-i\left[
 z+\frac{\gamma_B(z-\sigma H)}{\pi T_c}
 \sqrt{\Delta^2-z^2}\right]
 =-iE_{\rm eff,\sigma}(z).
\]
This connects Eq.~\eqref{eq:PhiF_thinF_supp} to Eq.~(3) of the main
text and Eq.~\eqref{eq:Eeff_supp} below. In the anomalous-block
convention used here, the retarded Green functions are
\endgroup
\begin{equation}
    G^R_\sigma(E)
    =
    \frac{
        E_{\rm eff,\sigma}(E)
    }{
        \sqrt{
            E_{\rm eff,\sigma}^2(E)-\Delta^2
        }
    },
    \label{eq:GR_eff_supp}
\end{equation}
and
\begin{equation}
    F^R_\sigma(E)
    =
    -\frac{i\Delta}
    {
        \sqrt{
            E_{\rm eff,\sigma}^2(E)-\Delta^2
        }
    } .
    \label{eq:FR_eff_supp}
\end{equation}
Here the effective energy is
\begin{equation}
    E_{\rm eff,\sigma}(E)
    =
    E+i\Gamma
    +
    \frac{
        \gamma_B
        \left(E+i\Gamma-\sigma H\right)
    }{
        \pi T_c
    }
    \sqrt{
        \Delta^2-\left(E+i\Gamma\right)^2
    } .
    \label{eq:Eeff_supp}
\end{equation}
\begingroup
The roots are fixed by analytic continuation from positive Matsubara
frequencies, rather than chosen independently as principal roots.
Writing $Q_\sigma^2=E_{\rm eff,\sigma}^2-\Delta^2$, the continued
branch satisfies
$Q_\sigma(i\omega)=i\sqrt{W_\sigma(\omega)^2+\Delta^2}$.
The auxiliary Matsubara function
$f_\sigma=\Delta/\sqrt{W_\sigma^2+\Delta^2}$ continues to
$f_\sigma^{\rm an}=i\Delta/Q_\sigma$; our anomalous-block convention is
$F_\sigma^R=-f_\sigma^{\rm an}$, consistently with
Eq.~\eqref{eq:FR_eff_supp} and $a_\sigma=iF_\sigma^R/(1+G_\sigma^R)$.
The Dynes parameter \(\Gamma\) is included by replacing \(0^+\) with
\(\Gamma>0\) consistently in both occurrences of $z$. It regularizes
the retarded functions and is not determined by the thin-F reduction.
The formal retarded prescription is recovered as \(\Gamma\to0^+\).

\endgroup
The spin-resolved density of states is
\begin{equation}
    N_\sigma(E)
    =
    N_0\,{\rm Re}\,G^R_\sigma(E).
    \label{eq:DOS_supp}
\end{equation}
Equations~\eqref{eq:GR_eff_supp}--\eqref{eq:DOS_supp} are the
spin-resolved spectral functions used in the main text.

It is useful to emphasize the physical content of
Eq.~\eqref{eq:Eeff_supp}. The exchange field does not merely shift the
BCS energy argument \(E\to E-\sigma H\). Instead, it enters through the
proximity-renormalized effective energy \(E_{\rm eff,\sigma}(E)\). As a
result, the SF density of states develops an additional
exchange-induced spectral singularity. For the parameter range used in
the MAR calculations, the relevant shifted singularity of \(N_+(E)\)
lies at negative energy; in the main text its position is denoted by
\(E_{\rm peak}<0\), and \(E_p=|E_{\rm peak}|\). The same singular
structure enters the MAR recurrence through \(a_\sigma(E)\) and is
therefore probed by the coherent MAR ladder.

The same reconstructed density of states controls the Josephson
anomalies of SFIFS junctions discussed in
Ref.~\cite{supp:Fominov2002}. There, in the antiparallel configuration,
the critical current is enhanced when the exchange-induced spectral
singularity is shifted toward the Fermi level, while in the parallel
configuration it is associated with the \(0\)-\(\pi\) transition. In
the present problem the same proximity-reconstructed Green functions
enter the coherent MAR recurrence through the Andreev amplitudes
\begin{equation}
    a_\sigma(E)
    =
    \frac{
        iF^R_\sigma(E)
    }{
        1+G^R_\sigma(E)
    } .
    \label{eq:a_sigma_thinF_supp}
\end{equation}
Thus the shifted MAR anomalies obtained in the main text originate
from the internal spectral reconstruction of the SF electrodes, rather
than from a rigid Zeeman displacement of BCS gap edges.

\SuppSection{Numerical method}
\label{secS:numerical}

\begingroup\SuppSubsection{Current evaluation and transparency averaging}\endgroup
The numerical implementation follows the standard Averin--Bardas
recurrence with the replacement
\begin{equation}
a(E)\rightarrow a_\sigma(E,H).
\label{eqS:numerical_replacement}
\end{equation}
For each spin sector the recurrence is solved independently and currents are summed according to Eq.~\eqref{eq:physical_current_supp}.
\begingroup
The recurrence and closed current expression needed for the symmetric parallel calculation are
also collected in Appendix B of the main text. Numerical evaluation
and energy-unit conventions are given below in
Sec.~\ref{secS:evaluation_units}, which separates the raw-current
calculation from the additional display smoothing used for Fig.~S1
and the feature-position extraction in Fig.~S3.\endgroup{}

\begingroup
A small Dynes parameter $\Gamma$ regularizes the retarded Green
functions. The energy cutoff, energy spacing, ladder extent, voltage
mesh, and differentiation window control different numerical errors
and must be distinguished. The figure-specific settings and the
separation between raw-current evaluation and differentiation are stated in Sec.~\ref{secS:evaluation_units}. 
\endgroup{}

For a diffusive nanobridge the final current is obtained by averaging the single-channel MAR current over the distribution of normal-state transparencies. Since the MAR recurrence gives the current for a fixed channel transparency \(D\), the averaged current must be calculated before performing the voltage derivative. In particular, the differential resistance shown in the main text is obtained as \begin{equation} \frac{dV}{dI} = \left[ \frac{d I_{\rm av}(V)}{dV} \right]^{-1}, \end{equation} where \(I_{\rm av}(V)\) denotes the transparency-averaged current. 

For a short diffusive constriction the Dorokhov distribution is 
~\cite{supp:dorokhov1984}

\begin{equation} \rho(D) = \frac{R_Q}{2R_N} \frac{1}{D\sqrt{1-D}}, \end{equation} 
\begingroup
where $R_Q=h/(2e^2)$ is the resistance quantum and $R_N$ in this
distribution is the total normal resistance of the multichannel
connector. Distinguish the physical single-channel current $I(V;D)$
from $j_D(v)=eR_N(D)I(V;D)/\Delta$, with $v=eV/\Delta$ and
$R_N(D)=R_Q/D$. The conductance-weighted average is
\begin{equation}
 j_{\rm av}(v)=
 \frac{\displaystyle\int_0^1dD\,\rho(D)D\,j_D(v)}
      {\displaystyle\int_0^1dD\,\rho(D)D}.
\end{equation}
It gives the physical current in Eq.~(S55) as
$I_{\rm av}(V)=\Delta j_{\rm av}(v)/(eR_N)$.
The substitution $u=\sqrt{1-D}$ yields
$j_{\rm av}(v)=\int_0^1du\,j_{1-u^2}(v)$.
Thus neither the resistances nor channel-normalized currents with an
unweighted mode density are averaged. The explicit channel sum,
quadrature form, and derivative convention are given in Appendix B of
the main text.\endgroup{}

\begingroup
For Fig.~3 of the main text, the differential resistance is formed
from the transparency-averaged current. The quadrature implements the
physical average above; the selected single-channel checks reported below
are not error bounds for this transparency-averaged curve.
\endgroup{}

\begingroup
\SuppSubsection{Numerical evaluation and energy units}
\label{secS:evaluation_units}
The retarded amplitudes are evaluated with $z=E+i\Gamma$, using the
branches specified in Appendix A of the main text. The transport
curves in Figs.~3 and 4 of the main text use
$\Gamma=0.005\Delta$. The unit-source amplitudes are
$\mathcal A=A/J_\sigma$ and $\mathcal B=B/J_\sigma$, with the
zero-source case understood by continuity. A complex tridiagonal solve
for $\mathcal B$ is followed by forward substitution for $\mathcal A$.
The energy cutoff and recurrence extent are separate controls.

Choosing $\Delta$ as the energy unit does not eliminate the material
ratio $\Delta/(\pi T_c)$. The spectrum depends on
$\gamma_B\Delta/(\pi T_c)$ and on the exchange parameter
\begin{equation}
 \eta=\frac{\gamma_B H}{\pi T_c}.
 \label{eq:dimensionless_spectral_supp}
\end{equation}
The compact label $\gamma_BH$ denotes $\eta$.
Spectral and thermal broadening are specified by $\Gamma/\Delta$ and
$T/\Delta$, respectively. Changing the thermal occupation weight alone
does not determine the parent gap $\Delta(T)$.

The full-range \emph{SFcFS} currents underlying Fig.~S1 use
$D=0.7$, $\gamma_B=0.01$, $\eta=0.3,0.5$,
$\Gamma/\Delta=0.005$, and $T=0$.
The derivative is extracted from a local polynomial fit,
\begin{equation}
 j(v)\simeq\sum_{m=0}^{M}c_m(v-v_0)^m,\qquad
 \left.\frac{dj}{dv}\right|_{v_0}=c_1.
 \label{eq:local_polynomial_supp}
\end{equation}
For Fig.~S1, the final extraction uses $M=3$, a nominal half-window
$0.035$ in $v$, at least seven points, and smoothly decreasing
weights toward the window edges. At voltages away from the conventional
MAR thresholds, the fitting set is restricted to the interval between
adjacent thresholds. 

For display in Fig.~S1, shape-preserving piecewise-cubic interpolation
places the curves on $2200$ uniformly spaced voltages. A cubic
Savitzky--Golay filter is then applied to the SFcFS curves with windows
of $15$ and $17$ display points for $\eta=0.3$ and $0.5$, respectively.
Fig.~S1 supplies
full-range context rather than precision extraction of fine structure.
\endgroup

\SuppSection{Additional figures}
\label{secS:additional_figures}

The following supplementary figures provide numerical checks and
additional context for the main-text results. 

Figure~S1 shows the
full-range differential-resistance spectra for the ScS and SFcFS
single-channel contacts. 

Figure~S2 illustrates the formal spin-sector
decomposition of the SFcFS current and verifies the symmetry under
bias reversal.

\begingroup
Figure~S3 compares the selected shifted MAR features with the approximate
spectral-interval scale
\[
    eV_n^{(p,\mathrm{calc})}/\Delta
    \simeq
    \frac{1+|E_{\rm peak}|/\Delta}{n}.
\]
The $n=4$ markers locate resolved maxima; the $n=3$ markers locate
maximum-slope points on the leading edges of overlapping structures.
Their field-dependent displacement is consistent with this scale, but
no exact relation between broadened resistance maxima and ideal spectral
thresholds is assumed. The spectral input, prediction-informed
selection windows, and scope of the comparison are specified in
Sec.~\ref{secS:evaluation_units}.
\endgroup

\begin{figure}[!htbp]
    \centering
    \includegraphics[width=0.7\textwidth]{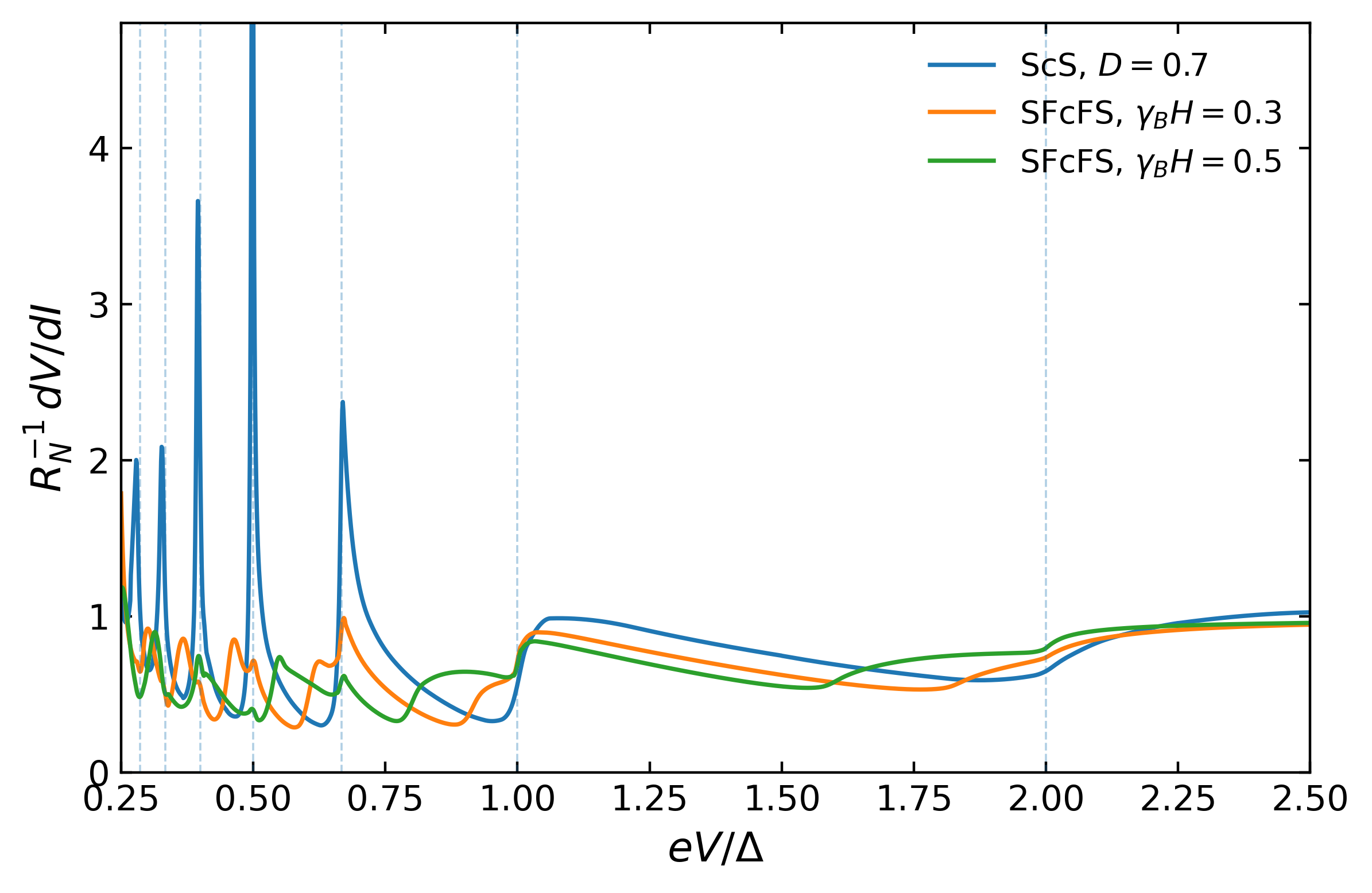}
    \caption{\begingroup
Full-range differential-resistance spectra for a conventional ScS
contact and an SFcFS contact with $D=0.7$. The SFcFS curves use
$\gamma_B=0.01$, $\Gamma=0.005\Delta$,
and the indicated values of $\eta$, denoted by $\gamma_BH$ in the
legend. The ordinate $R_N^{-1}dV/dI$ applies to these SFcFS curves.
Vertical dashed lines mark $eV/\Delta=2/n$.
The vertical scale is truncated. Interpolation and filtering are
described in Sec.~IV B. The figure provides full-range context for
Fig.~4 of the main text.
\endgroup}
    \label{fig:dVdIzoom}
\end{figure}

\begin{figure}[!htbp]
    \centering
    \includegraphics[width=0.6\textwidth]{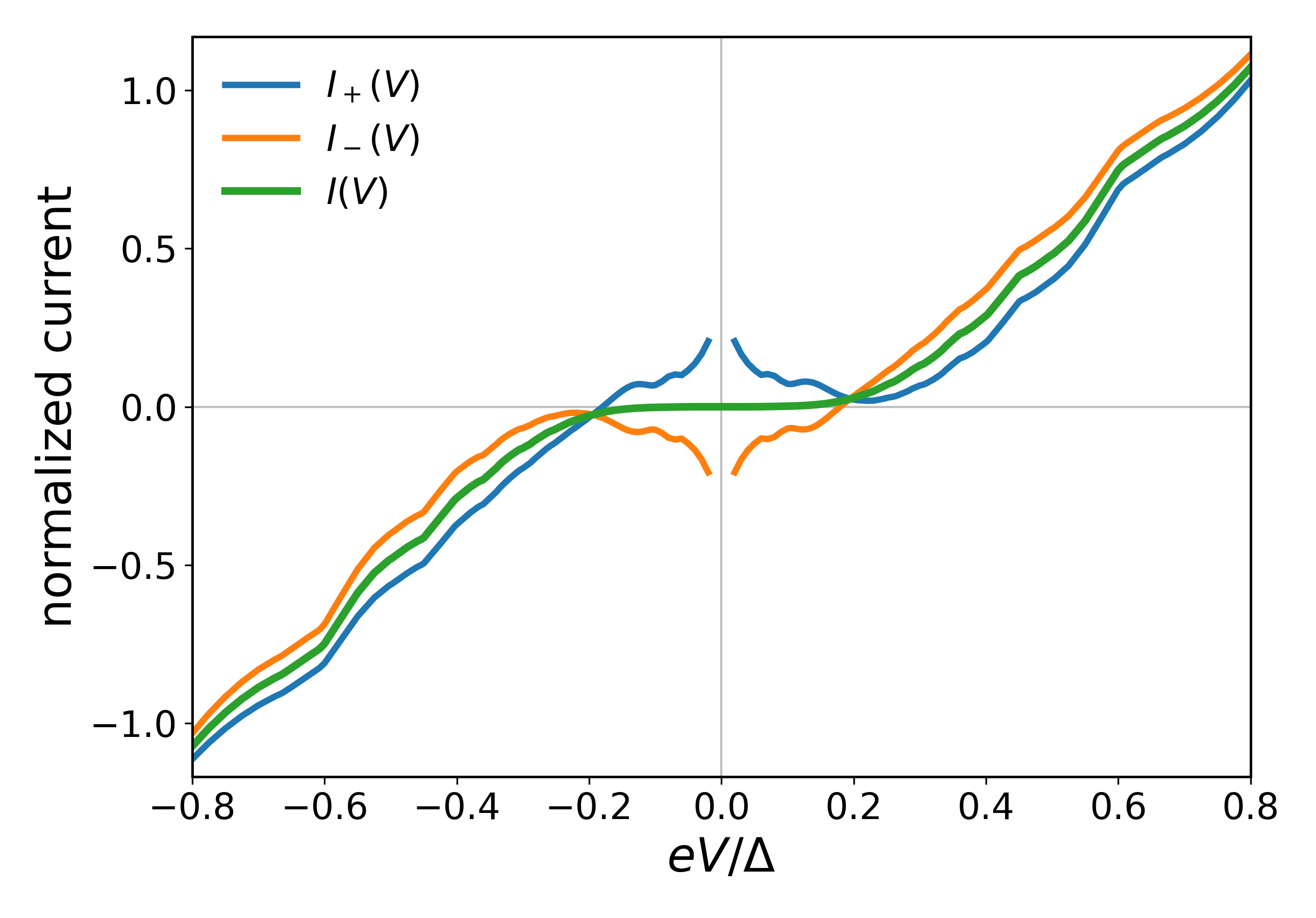}
\caption{
Spin-sector decomposition of the SFcFS current for \(D=0.7\),
\(\gamma_B=0.01\), and \(\gamma_BH=0.3\). The quantities
\(I_+(V)\) and \(I_-(V)\) are the sector-resolved contributions
associated with the two decoupled Nambu--spin sectors of the parallel
junction. They are not separately measurable charge currents.
Bias reversal interchanges the two sectors according to
\(I_+(-V,H)=-I_-(V,H)\), so that the physical current
\(I(V)=\frac{1}{2}\bigl[I_+(V)+I_-(V)\bigr]\) is antisymmetric
in \(V\). The figure provides a numerical check of
the spin-sector decomposition used in the main text.
}
    \label{fig:spin-resolved}
\end{figure}

\clearpage
\begin{figure}[!htbp]
\centering
\includegraphics[width=0.6\linewidth]{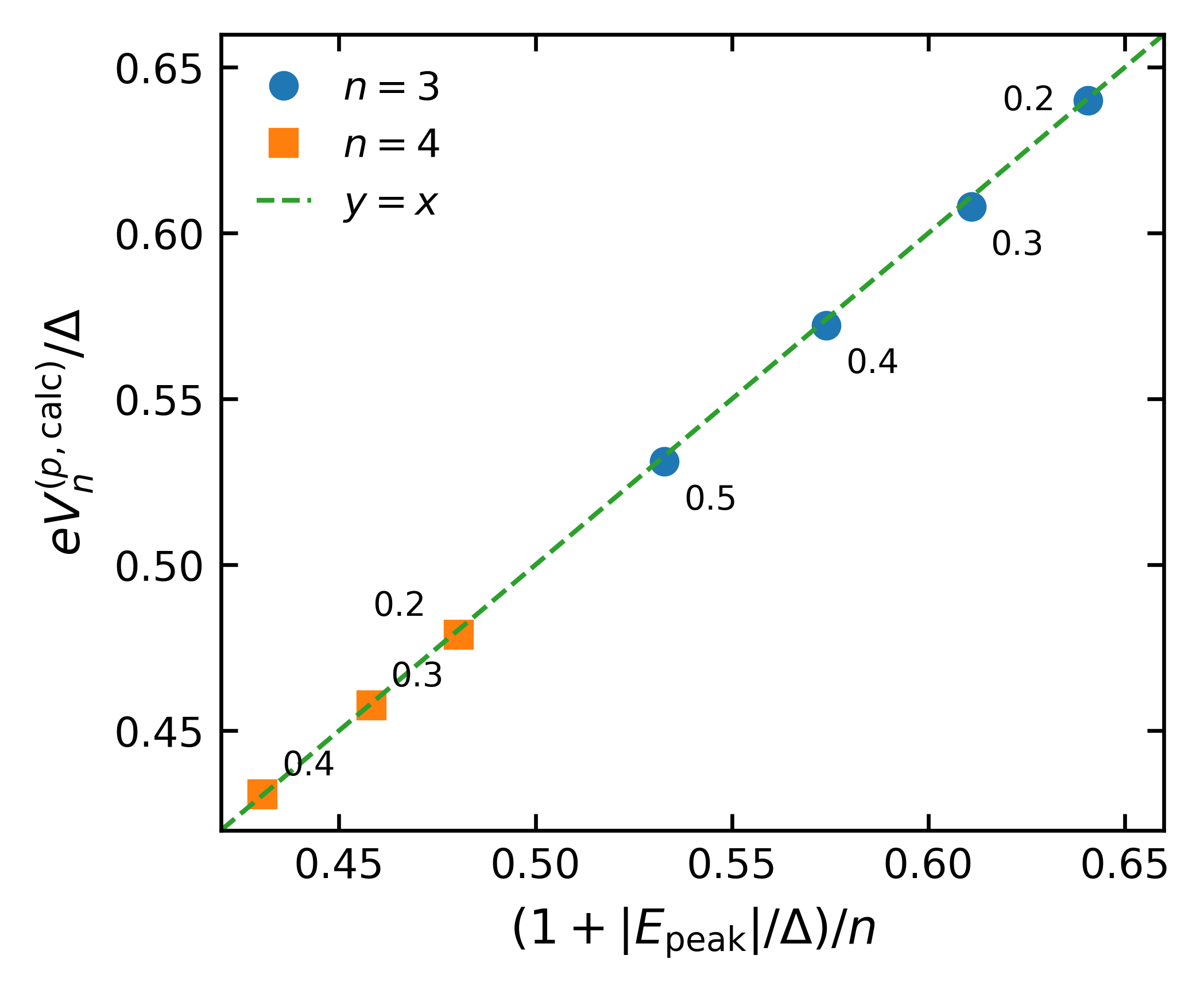}
\caption{\begingroup
Approximate spectral-interval comparison for the shifted MAR features.
The vertical axis shows the selected positions
$eV_n^{(p,\mathrm{calc})}/\Delta$ from the $dV/dI$ curves used for
Fig.~4 of the main text. The horizontal axis shows
$(1+|E_{\rm peak}|/\Delta)/n$, with the $N_+(E)$ peak evaluated at
$\gamma_B=0.01$
and $\Gamma/\Delta=0.005$.
Circles denote $n=3$, squares $n=4$; adjacent numbers give
$\eta=\gamma_BH/(\pi T_c)$. The dashed line is $y=x$.
For $n=4$, the shifted component is a resolved local maximum; for
$n=3$, its leading edge is identified by the maximum slope below the
conventional $2\Delta/3$ feature. The selection windows are centered
on the spectral estimate. This comparison illustrates field-dependent
shifts, not an independent precision test of peak positions or an
integration-error estimate. The extraction conventions and scope are
specified in Sec.~IV B.
\endgroup}
\label{fig:S_peak_tracking}
\end{figure}

\par